\documentclass[twocolumn,pra,aps,floatfix,footinbib,superscriptaddress]{revtex4-1}

\newmuskip\pFqmuskip

\newcommand*\pFq[6][8]{%
  \begingroup 
  \pFqmuskip=#1mu\relax
  \mathchardef\normalcomma=\mathcode`,
  \mathcode`\,=\string"8000
  \begingroup\lccode`\~=`\,
  \lowercase{\endgroup\let~}\pFqcomma
  {}_{#2}\mathrm{F}_{#3}{\left[\genfrac..{0pt}{}{#4}{#5};#6\right]}%
  \endgroup
}
\newcommand{\pFqcomma}{{\normalcomma}\mskip\pFqmuskip}

\usepackage{amsmath}
\usepackage{amssymb}
\usepackage{bigints}
\usepackage{bm}
\usepackage{comment}
\usepackage{etoolbox}
\usepackage{feynmf}
\usepackage{graphicx}
\usepackage{lipsum}
\usepackage{tikz}
\usepackage{verbatim}
\usepackage{xcolor}

\usepackage[export]{adjustbox}
\usepackage[colorlinks,urlcolor=blue,citecolor=blue,linkcolor=blue]{hyperref}
\usepackage{cleveref}
\usepackage[utf8]{inputenc}
\usepackage{pagecolor}
\usepackage[normalem]{ulem}

\usetikzlibrary{shapes}

\newcommand{\A}{\mathbf{A}}
\newcommand{\R}{\mathbf{R}}

\newcommand{\p}{\mathbf{p}}

\newcommand{\nn}{\nonumber}

\newcommand{\bra}[1]{\langle\left.\hspace{-0.05em}{#1}\hspace{+0.05em}\right|}
\newcommand{\ket}[1]{\left|\hspace{+0.05em}{#1}\hspace{-0.05em}\right.\rangle}

\renewcommand{\r}{\mathbf{r}}

\begin{document}

\title{Rydberg Exciton-Polaritons in a Magnetic Field}

\author{Emma~Laird}
\affiliation{School of Physics and Astronomy, Monash University, Victoria 3800, Australia}
\affiliation{ARC Center of Excellence in Future Low-Energy Electronics and Technologies, Monash University, Victoria 3800, Australia}

\author{Francesca M. Marchetti}
\affiliation{Departamento de F\'isica Te\'orica de la Materia Condensada \& Condensed Matter Physics Center (IFIMAC), Universidad Aut\'onoma de Madrid, Madrid 28049, Spain\\}

\author{Dmitry K.~Efimkin}
\affiliation{School of Physics and Astronomy, Monash University, Victoria 3800, Australia}
\affiliation{ARC Center of Excellence in Future Low-Energy Electronics and Technologies, Monash University, Victoria 3800, Australia}

\author{Meera M. Parish}
\affiliation{School of Physics and Astronomy, Monash University, Victoria 3800, Australia}
\affiliation{ARC Center of Excellence in Future Low-Energy Electronics and Technologies, Monash University, Victoria 3800, Australia}

\author{Jesper~Levinsen}
\affiliation{School of Physics and Astronomy, Monash University, Victoria 3800, Australia}
\affiliation{ARC Center of Excellence in Future Low-Energy Electronics and Technologies, Monash University, Victoria 3800, Australia}

\date{\today}

\begin{abstract}

We theoretically investigate exciton-polaritons in a two-dimensional (2D) semiconductor heterostructure, where a static magnetic field is applied perpendicular to the plane. To explore the interplay between magnetic field and a strong light-matter coupling, we employ a fully microscopic theory that explicitly incorporates electrons, holes and photons in a semiconductor microcavity. Furthermore, we exploit a mapping between the 2D harmonic oscillator and the 2D hydrogen atom that allows us to efficiently solve the problem numerically for the entire Rydberg series as well as for the ground-state exciton.  In contrast to previous approaches, we can readily obtain the real-space exciton wave functions and we show how they shrink in size with increasing magnetic field, which mirrors their increasing interaction energy and oscillator strength.  We compare our theory with recent experiments on exciton-polaritons in GaAs heterostructures in an external magnetic field and we find excellent agreement with the measured polariton energies. Crucially, we are able to capture the observed light-induced changes to the exciton in the regime of very strong light-matter coupling where a perturbative coupled oscillator description breaks down.  Our work can guide future experimental efforts to engineer and control Rydberg excitons and exciton-polaritons in a range of 2D materials.

\end{abstract}

\pacs{}

\maketitle


\section{Introduction}
\label{sec:Introduction}

Excitons are bound electron-hole pairs that can be optically addressed and manipulated in a direct bandgap semiconductor.  While the lowest energy ground-state exciton is the easiest to access experimentally --- it has the largest binding energy and the strongest coupling to light --- the excited Rydberg excitons have recently been gaining much attention due to their strong dipole-dipole interactions~\cite{Kazimierczuk2014,Heckotter2021}. In particular, by coupling the Rydberg states to photons in a microcavity, there is the possibility of creating Rydberg exciton-polaritons~\cite{Rypolariton,Gu2021,Orfanakis2022} with giant optical non-linearities, which have potential applications in quantum photonics~\cite{Walther2018}.  However, their reduced oscillator strength makes it challenging to achieve the strong light-matter coupling regime, which requires the coherent energy exchange between excitons and cavity photons to be faster than the rate of dissipation~\cite{Microcavities}.

One route to enhancing the coupling to light is to subject the excitons to a static magnetic field, since this effectively creates an additional confinement on the electron-hole wave function~\cite{Stafford1990}.  Experiments on GaAs heterostructures have already successfully tuned exciton-polaritons with a magnetic field and observed the emergence of strong coupling for Rydberg excitons at sufficiently large fields~\cite{Tignon1995,PhysRevB.91.075309,PhysRevB.96.081402}.  At the same time, the light-matter interactions themselves can modify the electron-hole wave function~\cite{Cortese2021}, leading to either a reduction or an enhancement of the exciton size (for lower or upper polaritons, respectively)~\cite{KHURGIN2001307,Citrin2003,PhysRevRes_1_033120}.  Indeed, light-induced changes to the exciton radius were recently demonstrated in multiple GaAs quantum wells embedded into a microcavity~\cite{PhysRevLett_119_027401}. Here, a magnetic field was used to probe the diamagnetic shift of polaritons in the very strong coupling regime, where the ratio between the exciton-photon coupling strength to the exciton binding energy approaches unity.  However, as yet, there is no theory for the exciton-polariton that can describe the interplay between a very strong light-matter coupling and the effect of a magnetic field.

In this paper, we present an exact microscopic calculation of a two-dimensional (2D) exciton-polariton in a static perpendicular magnetic field. Our work relies on two key innovations: a recently developed microscopic model for the polariton in the absence of a magnetic field, which exactly incorporates the light-induced modification of the electron-hole pair in a semiconductor microcavity \cite{PhysRevRes_1_033120}; and an exact mapping between the 2D harmonic oscillator and the 2D hydrogen atom. The latter allows us to solve the problem in momentum space, thus giving us easy access to the spectrum of weakly bound Rydberg states as well as the ground-state exciton. We find excellent agreement between our theory and the recent experiments on polaritons in a magnetic field~\cite{PhysRevLett_119_027401,PhysRevB.96.081402}, thereby illustrating the accuracy and versatility of our approach. In particular, we provide a direct and quantitative comparison with experiments in the very strong light-matter coupling regime~\cite{PhysRevLett_119_027401}, where a simple coupled oscillator description fails. Ultimately, we expect our theory to be able to guide experiments that seek to control excitons with light and magnetic fields, for instance in atomically thin materials such as transition-metal dichalcogenides (TMDs).

This paper is organized as follows: Our model for an exciton-polariton in a magnetic field is introduced in Sec.~\ref{sec:Model}.  In Sec.~\ref{sec:Exciton_Problem}, we first solve the exciton problem in the absence of light-matter coupling and determine the properties of the Rydberg series in a magnetic field.  Then in Sec.~\ref{sec:Exciton-Polariton_Problem}, we include a strong light-matter coupling, properly accounting for the renormalization of the cavity photon~\cite{PhysRevRes_1_033120}.  To illustrate the power and utility of our approach, we compare our theory with recent experimental measurements on exciton-polaritons in magnetic fields~\cite{PhysRevB.96.081402,PhysRevLett_119_027401} in Sec.~\ref{sec:Theoretical_Comparison_to_Experiment}.  Finally, we indicate some future directions of the research in Sec.~\ref{sec:Conclusions_and_Outlook}.

\section{Model}
\label{sec:Model}

In this work, we consider a 2D semiconductor embedded in an optical microcavity and subjected to a static perpendicular magnetic field.  The Hamiltonian can be decomposed into the key elements of this system, corresponding to the matter excitations of the semiconductor, the light field within the microcavity, and the light-matter coupling:
\begin{align}
\label{eq:Hamiltonian}
\hat{H}=
\hat{H}_\mathrm{mat}+
\hat{H}_\mathrm{ph}+
\hat{H}_{\mathrm{ph}\text{-}\mathrm{mat}}\,.
\end{align}

When a photon of the microcavity is incident on the semiconductor with an energy that exceeds the bandgap, a negatively charged electron is excited into the conduction band, leaving behind a positively charged hole in the valence band.  To describe such an electron-hole pair, we can write the matter part of the Hamiltonian in first quantization as follows:
\begin{align}
\label{eq:H_mat}
\hat{H}_{\mathrm{mat}}=\,\,&\frac1{2m_e}\!\left[\hspace{0.1em}\hat{\p}_e+\frac{e}{c}\A(\hat{\r}_e)\right]^2\!+\frac1{2m_h}\!\left[\hspace{0.1em}\hat{\p}_h-\frac{e}{c}\A(\hat{\r}_h)\right]^2\!\nn\\&\hspace{-0.6mm}+V(|\hat{\r}_e-\hat{\r}_h|)\,,
\end{align}
where, for simplicity, we neglect the spin degrees of freedom, and we assume we are close to the band edge where the conduction and valence bands have parabolic dispersions.  Here, $\hat{\r}_e$ and $\hat{\r}_h$ ($\hat{\p}_e$ and $\hat{\p}_h$) are the in-plane position (momentum) operators for an electron and a hole respectively, while $m_e$ and $m_h$ are their effective masses, $e$ is the elementary charge (i.e., the magnitude of the electron's charge), and $c$ is the speed of light.  For convenience, we choose to work in the symmetric gauge such that the vector potential is given by $\A(\r)=\frac12(\mathbf{B}\times\r)=(-B y/2,\,B x/2,\,0)$, where $\mathbf{B}=\nabla\times\A(\r)=(0,\,0,\,B)$ corresponds to a magnetic field of strength $B$ applied perpendicular to the plane.  Such a gauge is also consistent with the Coulomb gauge $[\nabla\cdot\A(\r) = 0]$ which is used to describe the coupling to light.

We employ the 2D Coulomb potential to characterize the electron-hole interactions:
\begin{align}
\label{eq:Coulomb_potential}
V^\mathrm{C}(|\r_e-\r_h|)=-\frac{e^2}{\varepsilon|\r_e-\r_h|}\,,
\end{align}
where $\varepsilon$ is the static dielectric constant of the semiconductor.  However, as we shall discuss, our approach is valid for other interaction potentials, including the Rytova--Keldysh one~\cite{Rytova,Keldysh,PhysRevB_84_085406} which properly models the dielectric screening in atomically thin TMDs~\cite{PhysRevB_89_205436,PhysRevLett_113_076802}.  Notice that we use Gaussian units ($4\pi\varepsilon_0=1$), we measure energies with respect to the bandgap, and we set both $\hbar$ and the system area to unity.

In the single-particle (e.g., electron) problem without a magnetic field, the momentum operator $\hat{\mathbf{p}}_e$ commutes with the Hamiltonian and acts as the generator of translations.  However, in the presence of a uniform magnetic field $\mathbf{B}$, while the system remains translationally invariant, the correct generator of translations becomes the aptly named magnetic momentum, which in the symmetric gauge is given by $\hat{\mathbf{K}}_e=\hat{\p}_e-\frac{e}{c}\A(\hat{\r}_e)$~\cite{Quantum_Hall_Effect}.  Similarly, in the electron-hole problem, we have the total magnetic momentum operator
\begin{align}
\label{eq:pseudomomentum}
\hspace{-4.0mm}\hat{\mathbf{K}}=\hat{\p}_e-\frac{e}{c}\A(\hat{\r}_e)+\hat{\p}_h+\frac{e}{c}\A(\hat{\r}_h)=\hat{\mathbf{P}}-\frac{e}{2c}\mathbf{B}\times\hat{\r}\,,
\end{align}
where $\hat{\mathbf{P}}=\hat{\p}_e+\hat{\p}_h$ is the operator for the center-of-mass momentum and $\hat{\r}=\hat{\r}_e-\hat{\r}_h$ for the relative position.  Note that while the magnetic momentum of an electron-hole pair is a good quantum number of the matter Hamiltonian~\eqref{eq:H_mat}, the center-of-mass momentum is generally not.

For the photonic part of the Hamiltonian~\eqref{eq:Hamiltonian}, we use the second quantized form
\begin{align}
\label{eq:H_ph}
\hat{H}_\mathrm{ph}=\omega\hspace{0.25mm}\hat{c}^\dagger\hat{c}\,,
\end{align}
where the operator $\hat{c}^\dagger$ ($\hspace{0.10mm}\hat{c}\hspace{0.25mm}$) creates (annihilates) a single cavity photon with bare frequency $\omega$. For simplicity, we only consider photons at normal incidence to the semiconducting plane, i.e., with zero in-plane momentum. However, even when the photon in-plane momentum is non-zero, it is still negligible compared to the typical momenta of the electrons and holes. Therefore, we can straightforwardly account for a non-zero photon momentum in a planar cavity via a shift of the cavity frequency.

We assume that the length scale over which light interacts with matter is much shorter than the magnetic length or any other relevant length scale in the system.  That is, we assume that a photon creates an electron and a hole at essentially zero relative separation. Thus, within the dipole approximation, the light-matter coupling term can be written as
\begin{align}
\label{eq:H_ph-mat}
\hat{H}_{\mathrm{ph}\text{-}\mathrm{mat}}=g\hspace{0.25mm}\delta^2(\hat\r)\!\left[\hat{c}^\dagger+\hat{c}\hspace{0.28mm}\right],
\end{align}
where $\delta^2(\r)$ is the 2D Dirac delta function and the constant $g$ describes the strength of the short-range contact interaction. Since the normally incident photon creates an electron-hole pair with both zero centre-of-mass momentum and zero electron-hole separation, the magnetic momentum $\mathbf{K}$ is also zero according to Eq.~\eqref{eq:pseudomomentum}. Therefore, because $\mathbf{K}$ is a good quantum number, we can set it to zero throughout the entire problem.

We now transform the full light-matter coupled system~\eqref{eq:Hamiltonian} into the center-of-mass frame of the electron-hole pair:
\begin{subequations}
\label{eq:transformation}
\begin{align}
\label{eq:transformation_A}
\hat{H}'&=\hat{U}\hat{H}\hspace{+0.1em}\hat{U}^\dagger\,,\\
\label{eq:transformation_B}
\hat{U}&=\mathrm{exp}\!\left\{-i\hspace{+0.1em}\hat{\mathbf{P}}\cdot\hat{\R}\right\}=\mathrm{exp}\!\left\{-i\hspace{-0.1em}\left[\hat{\mathbf{K}}+\frac{e}{2c}(\mathbf{B}\times\hat{\r})\right]\!\cdot\hat{\R}\right\},
\end{align}
\end{subequations}
where $\hat{\R}=(m_e\hat{\r}_e+m_h\hat{\r}_h)/(m_e+m_h)$ is the operator for the pair's center-of-mass position~\footnote{This transformation to the center-of-mass frame for two bodies under the influence of a magnetic field was first performed by Lamb~\cite{Lamb} in his dealings with the hydrogen atom, and subsequently performed in the context of excitons by Gor’kov and Dzyaloshinskii~\cite{Gorkov&Dzyaloshinskii}.}.  We can immediately see that the operator~\eqref{eq:transformation_B} of this unitary transformation commutes with both $\hat{H}_\mathrm{ph}$~\eqref{eq:H_ph} and $\hat{H}_{\mathrm{ph}\text{-}\mathrm{mat}}$~\eqref{eq:H_ph-mat}, such that they are left unchanged.  The only effect of Eq.~\eqref{eq:transformation_A} is to recast $\hat{H}_\mathrm{mat}$~\eqref{eq:H_mat} in terms of the positions and momenta of the pair's center-of-mass and relative motions --- in a way that respects the translational symmetry of the system in a magnetic field.  After some algebra, we arrive at the transformed matter part of the Hamiltonian:
\begin{align}
\label{eq:new_H_mat}
\hat{H}_{\mathrm{mat}}'=\,\,&\frac{\hat{\p}^2}{2\mu}+\frac{e\eta}{2\mu c}\mathbf{B}\cdot(\hat{\r}\times\hat{\p})+\frac{e^2}{8\mu c^2}(\mathbf{B}\times\hat{\r})^2\nn\\&\hspace{-0.6mm}+V(|\hat{\r}|)+\frac{e}{M\hspace{-0.1em}c}(\hat{\mathbf{K}}\times\mathbf{B})\cdot\hat{\r}+\frac{\hat{\mathbf{K}}^2}{2M}\,,
\end{align}
with
\begin{align}
\label{eq:mass_factors}
\hspace{-4.0mm}\mu=\frac{m_em_h}{m_e+m_h}\,,\quad M=m_e+m_h\,,\quad\eta=\frac{m_h-m_e}{m_h+m_e}\,,
\end{align}
and where $\hat{\p}=(m_h\hat{\p}_e-m_e\hat{\p}_h)/(m_e+m_h)$ is the opera-\linebreak tor for the pair's relative momentum~\footnote{In this form, one can show that $[\hat{H}_{\mathrm{mat}},\hat{\mathbf{K}}]=[\hat{H}_{\mathrm{mat}}',\hat{\mathbf{K}}]=0$, since $[\hat{\mathbf{K}},\hat{\mathbf{P}}]=0$, while $[\hat{H}_{\mathrm{mat}},\hat{\mathbf{P}}]\neq0$.}.

As mentioned above, the optically generated electron-hole pair has zero magnetic momentum and therefore the $\hat{\mathbf{K}}$-dependent terms can be ignored in Eq.~\eqref{eq:new_H_mat}. Furthermore, the second term on the right-hand side can be re-expressed as
\begin{align}
\label{eq:2nd_term}
\frac{e\eta}{2\mu c}\mathbf{B}\cdot(\hat{\r}\times\hat{\p})\equiv\frac{\omega_{c,\hspace{0.05em}e}-\omega_{c,\hspace{0.05em}h}}{2}\hat{L}_z\,,
\end{align}
where $\omega_{c,\hspace{0.05em}j}=eB/(m_{j}c)$ [$j=e,h$] are the cyclotron frequencies of the electron and hole, and $\hat{L}_z=\hat{x}\hat{p}_y-\hat{y}\hat{p}_x$ is the $z$ component of the orbital angular momentum operator for their relative motion.  Crucially, the only states of matter that couple to light in our model~\eqref{eq:H_ph-mat} are the isotropic $s$ excitons, and $L_z$ is zero for these states.  Hence, the ``orbital'' term~\eqref{eq:2nd_term} above can also be discarded~\footnote{We mention furthermore that this ``orbital'' term is very small for the case of monolayer TMDs where $m_e\simeq m_h$.}, leaving us with the following effective matter part of the Hamiltonian:
\begin{align}
\label{eq:final_H_mat}
\hat{H}_{\mathrm{mat}}^{\prime\,\mathrm{eff}}=\frac{\hat{\p}^2}{2\mu}+
\frac12\mu\omega_c^2
|\hat{\r}|^2+V(|\hat{\r}|)\,,
\end{align}
where we introduce the electron-hole cyclotron frequency $\omega_c=eB/(2\mu c)$~\footnote{Of course, the operator $\hat{\p}^2\hspace{-0.4mm}/(2\mu)$ also contains $\hat{L}_z$, but this can only be separated out by projecting onto the position basis and using polar co-ordinates, as we do in the next section~(\ref{sec:Exciton_Problem}).}.  This reduces to the electron cyclo-\linebreak tron frequency $\omega_{c,\hspace{0.05em}e}$ when the electron and hole masses are equal.

Finally, we can write down the full effective Hamiltonian for an exciton-polariton in a magnetic field,
\begin{align}
\label{eq:effective_Hamiltonian}
\hat{H}_\mathrm{eff}'=
\hat{H}_{\mathrm{mat}}^{\prime\,\mathrm{eff}}+
\hat{H}_\mathrm{ph}+
\hat{H}_{\mathrm{ph}\text{-}\mathrm{mat}}\,,
\end{align}
which is used in the ensuing sections.

\section{Exciton in a magnetic field}
\label{sec:Exciton_Problem}

In the current section, we revisit the scenario of a two-dimensional exciton in a static perpendicular magnetic field, without any light-matter coupling, and we provide a new and powerful procedure to efficiently solve the problem numerically.  Previously, this problem has been addressed using a variety of theoretical approaches, including the WKB method~\cite{JPhysSocJapan_22_181}, variational wave functions~\cite{Rosner_1984,Feng_thesis}, and a two-point Pad{\'e} approximant that interpolates between the perturbative results in the weak- and strong-field limits~\cite{PhysRevB_33_8336}.  Here, we exploit a real-space mapping between the hydrogen problem and the harmonic oscillator, allowing us to convert the Schr\"{o}dinger equation into a form that is readily solved by modified quadrature integration methods.  This strategy is computationally cheaper than numerically integrating the Schr{\"o}dinger equation~\cite{JPhysSocJapan_29_1258,Whittaker_&_Elliott,PhysRevB_37_2759,PhysRevB_39_7697} and yields the fullest solution --- numerically converged energies and wave functions for the ground state and at least a dozen excited Rydberg states.  Our results are directly relevant to thin\linebreak quantum-well semiconductors and can also be extended to apply to current experiments on atomically thin materials~\cite{Stier2016,Stier2018} with only minor modifications. In particular, we expect the excited Rydberg states in TMDs to be well described by an unscreened Coulomb potential, since dielectric screening only affects the Rytova--Keldysh potential~\cite{Rytova,Keldysh} at small electron-hole separation.

Specializing to the optically active $s$ series of excitons, we begin by writing down the real-space representation of the Schr\"{o}dinger equation,
\begin{align}
\label{eq:exc_s-wave_with-B_Ch7}
E\varphi(r)=\left[-\frac1{2\mu}\!\left(\frac{d^2}{dr^2}+\frac1r\frac{d}{dr}\right)\!+\frac{\mu\omega_c^2}2r^2-\frac{e^2}\varepsilon\frac1r\right]\!\varphi(r)\,,
\end{align}
with $r\equiv|\r|$ and where we make use of the effective matter Hamiltonian $\hat{H}_{\mathrm{mat}}^{\prime\,\mathrm{eff}}$ in Eq.~\eqref{eq:final_H_mat}.  The energy $E$ is measured relative to the bandgap of the semiconductor, and $\varphi(r)$ is the wave function for the relative motion of the electron and hole (in the rest frame of the magnetic momentum).

Rather than proceeding to solve the above differential equation in real space, which is the usual approach, we Fourier transform the system into an integral equation in momentum space.  The only complication arising from such a tactic is that the magnetic-field term converts to a derivative in the momentum representation, since it is a function of $r^2$.  To circumvent this issue, we first rescale the real-space variable as $r\to\rho=r^2/(8a_0^2)$ and we introduce the dimensionless wave function $\widetilde\varphi(\rho)=a_0\varphi(r)$.  The Schr\"{o}dinger equation~\eqref{eq:exc_s-wave_with-B_Ch7} hence becomes
\begin{align}
\label{eq:exc_rho-eq_Ch7}
\frac{2\hspace{0.01em}\bar{E}}\rho\hspace{0.1em}\widetilde\varphi(\rho)=\left[-\!\left(\frac{d^2}{d\rho^2}+\frac1\rho\frac{d}{d\rho}\right)\!+4\hspace{0.1em}\bar\omega_c^2-\frac1{\sqrt{2\hspace{0.01em}\rho^3}}\right]\!\widetilde\varphi(\rho)\,.
\end{align}
Here and subsequently, we employ an overbar to denote quantities that are expressed in terms of 2D Rydberg units, i.e., those that are rescaled by the exciton binding energy $R=2\mu e^4/\varepsilon^2=1/(2\mu a_0^2)$, where $a_0=\varepsilon/(2\mu e^2)$ is the exciton Bohr radius. In other words, $\bar{\omega}_c\equiv\omega_c/R$ and $\bar{E}\equiv E/R$.  Note that we have kept the same order of terms here in $\rho$ space as before in $r$ space.  Most interestingly, this change of variable takes advantage of a mapping in 2D between the harmonic oscillator and the hydrogen problem~\cite{Duru1982}, where the parameter $\bar{E}$ plays the role of the strength of the Coulomb interaction, and $\bar\omega_c^2$ the role of the energy.  The novelty in the present transformation is the additional Coulomb potential in Eq.~\eqref{eq:exc_s-wave_with-B_Ch7} which transforms into a potential $\propto1/\rho^{3/2}$ in Eq.~\eqref{eq:exc_rho-eq_Ch7} \footnote{Alternatively, it is also possible to use a similar transformation to map Eq.~\eqref{eq:exc_s-wave_with-B_Ch7} onto an anharmonic oscillator~\cite{Hoangdo2013}, after which it can be solved by an iterative procedure.}.

\begin{figure}[b]
\begin{center}
\hspace{-1mm}
\includegraphics[trim = 25mm 2mm 26mm 21mm, clip, scale = 0.45]{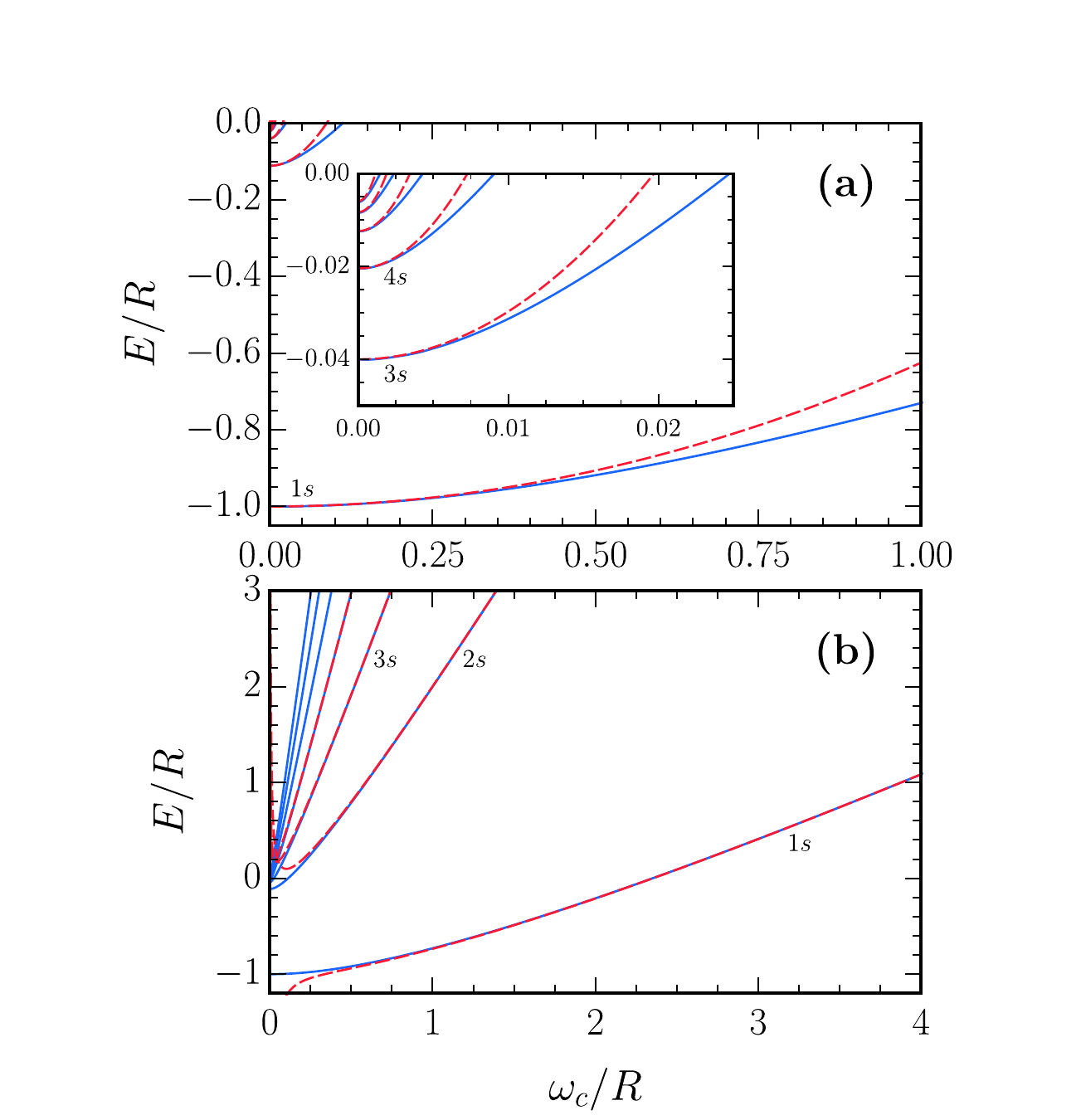}
\caption{Exciton energy as a function of magnetic field for the seven lowest energy $s$ states.  We focus on the low- and high-field limits respectively in the upper (a) and lower (b) pan-\linebreak els.  The solid blue lines correspond to the numerically exact results of our theory, Eq.~\eqref{eq:exciton_kappa}, while the dashed red lines are the results of perturbation theory for weak and strong magnetic fields, Eq.~\eqref{eq:weak-field_energies} in (a) and Eq.~\eqref{eq:strong-field_energies} in (b), respectively.}
\label{fig:fig_exciton_2}
\end{center}
\end{figure}

Now we Fourier transform the excitonic Schr\"{o}dinger equation into the rescaled momentum space by applying the operator $\int\!{d}^{\hspace{0.05em}2}\!\hspace{0.01em}\bm\rho\,{e}^{-i\bm\kappa\cdot\bm\rho}\{\,\cdot\,\}$ to Eq.~\eqref{eq:exc_rho-eq_Ch7}, where $\bm\kappa$ is the  conjugate variable to $\bm\rho$.  The result is given below, where $\Gamma(x)$ is the Euler gamma function and the summations are two-dimensional, $\sum_{\bm\kappa}\equiv\int\!{d}^{\hspace{0.05em}2}\!\hspace{0.10em}\bm\kappa/(2\pi)^2$:
\begin{align}
\label{eq:exciton_kappa}
\bar{E}\hspace{-0.4mm}\sum_{\bm\kappa'}\hspace{-0.5mm}\frac{4\pi\hspace{0.1em}\widetilde{\varphi}_{\kappa'}}{|\bm\kappa-\bm\kappa'|}=\kappa^2\hspace{0.1em}\widetilde{\varphi}_{\kappa}+4\hspace{0.1em}\bar{\omega}_c^2\hspace{0.1em}\widetilde{\varphi}_{\kappa}-\frac{\Gamma(1/4)}{\Gamma(3/4)}\hspace{-0.4mm}\sum_{\bm\kappa'}\hspace{-0.5mm}\frac{\pi\hspace{0.1em}\widetilde{\varphi}_{\kappa'}}{\sqrt{|\bm\kappa-\bm\kappa'|}}\,.
\end{align}
Again, we have kept the same order of terms here as in Eq.~\eqref{eq:exc_s-wave_with-B_Ch7}.

Equation~\eqref{eq:exciton_kappa} is a key result of our work, since it permits an efficient numerical solution of the problem of excitons in a magnetic field.  We can view the above expression as an eigenvalue-matrix problem, where the transformed wave function $\widetilde{\varphi}_{\kappa}$ is the eigenvector, the squared cyclotron frequency $\bar{\omega}_c^2$ is the eigenvalue, and the energy $\bar{E}$ is an input parameter.  These types of problems are re-\linebreak adily solved numerically by using, e.g., Gauss--Legendre quadrature methods~\cite{numerical_recipes}.  Importantly here, the two non-diagonal terms feature a pole at $\bm\kappa=\bm\kappa'$.  Rather than fictitiously removing the $\bm\kappa=\bm\kappa'$ elements from the two matrices, we implement a ``subtraction scheme'', which cancels out the singularities and significantly speeds up numerical convergence at high momenta.  The details are relegated to Appendix~\ref{sec:Appendix_A}.  This effectively means that we can calculate an entire spectrum (such as those presented in this paper), including highly excited Rydberg states, in less than a minute on a standard laptop.  Notice that although Eq.~\eqref{eq:exciton_kappa} specifically applies to the case of 2D Coulomb interactions, it can be adapted for other interaction potentials by modifying (only) the rightmost term --- we elaborate on this in Appendix~\ref{sec:Appendix_B}.

\subsection{Exciton energy spectrum}

The energy spectrum of a 2D exciton in the presence of a static perpendicular magnetic field --- as obtained by solving Eq.~\eqref{eq:exciton_kappa} --- is displayed in Fig.~\ref{fig:fig_exciton_2}.  Our numerically exact results are compared with the results of non-degenerate stationary-state perturbation theory valid at either low (a) or high (b) magnetic fields~\footnote{We can use a non-degenerate theory to calculate the perturbative corrections to the energies and wave functions at zero separation for the $s$ excitons.  At every order, in the low-field limit, this involves evaluating the matrix elements of $r^2$ in the hydrogen basis, and in the high-field limit, the matrix elements of $1/r$ in the harmonic oscillator basis.  Both functions of $r$ are rotationally symmetric.  Thus, if we were to calculate any such matrix element for an $s$ state and a higher angular momentum state, then we would obtain a radial integral multiplied by the angular integral $\int_0^{2\pi}\!d\theta\hspace{0.75mm}\mathrm{exp}(il\theta)$, where $l$ is a non-zero integer --- which always gives zero. In other words, these terms do not contribute to the perturbative expansions, and we can consider only the $s$ states which are non-degenerate.}.  The perturbative expressions were previously determined analytically in Ref.~\cite{PhysRevB_33_8336}, and we report them here for completeness.

The upper panel (a) of Fig.~\ref{fig:fig_exciton_2} illustrates the weak-field regime where the magnetic field can be treated as a perturbation.  The perturbative expansion for the $s$-exciton energies in this limit is given by
\begin{align}
\label{eq:weak-field_energies}
\bar{E}_{ns}\simeq&\hspace{1.1mm}\bar{E}_{ns}^{\mathrm{hyd}}+\bar{\omega}_c^2\times[5n(n-1)+3]\!\left(\frac{2n-1}{2\sqrt{2}}\right)^{\!\!2}\!+\mathcal{O}(\bar\omega_c^4)\,,
\end{align}
where $\bar{E}_{ns}^{\mathrm{hyd}}=-1/(2n-1)^2$ are the zero-field energies and $n=1,\,2,\,3,\,...$ is the principal (or radial) quantum number of the excitonic states~\cite{PhysRevB_33_8336}.  We can observe that for more highly excited states, the perturbative energies are accurate for an increasingly narrow range of magnetic fields.  In other words, as the field increases, higher excited states approach the strong-field limit more quickly than less excited states.

The lower panel (b) of Fig.~\ref{fig:fig_exciton_2} depicts the strong-field regime where the Coulomb potential can be treated as a perturbation.  The corresponding energy expansion is~\cite{PhysRevB_33_8336}
\begin{align}
\label{eq:strong-field_energies}
\bar{E}_{N\hspace{-0.2mm}s}\simeq&\hspace{1.2mm}2\hspace{0.1em}\bar{\omega}_c\!\left[N+\frac12+\alpha_N^{(1)}x+\alpha_N^{(2)}x^2+\alpha_N^{(3)}x^3+\alpha_N^{(4)}x^4\right]\nn\\&\hspace{-0.8mm}\hspace{1.2mm}+\mathcal{O}(x^5)\,,
\end{align}
where $x=\sqrt{\pi/(8\hspace{0.1em}\bar{\omega}_c)}\hspace{0.1em}$ and the numerical coefficients $\alpha_N^{(i)}$\linebreak are tabulated in a footnote~\footnote{The coefficients $\alpha_N^{(i)}$ in the perturbation expansion [Eq. \eqref{eq:strong-field_energies}] for the exciton energies at high magnetic fields are enumerated below:
\begin{center}
\begin{tabular}{ c c c c c }
\hspace{1.1mm} $N$ & $i=1$ & $i=2$ & $i=3$ & $i=4$ \\
\hspace{1.1mm} $0$ & $-1$ & $-0.44010149$ & $-0.2331170$ & $-0.07267451$ \\
\hspace{1.1mm} $1$ & $-3/4$ & $-0.11128950$ & $0.0339248$ & $0.0423287$ \\
\hspace{1.1mm} $2$ & $-41/64$ & $-0.052408254$ & $0.0271247$ & $0.0141383$ \\
\hspace{1.1mm} $3$ & $-147/256$ & $-0.031171820$ & $0.0192753$ & $0.0061608$
\end{tabular}
\end{center}
These values have been taken from Table~II of Ref.~\cite{PhysRevB_33_8336} and are accurate to the number of figures listed.}.  The index $N=n-1$ now designates the Landau level, and in Fig.~\ref{fig:fig_exciton_2} we have $N=0$ for the ground state, $N=1$ for the first excited state, and so on.  Consistent with and opposite to panel (a), we can see that the higher the energy level, the larger the magnetic-field range over which perturbation theory is accurate in panel (b).  Note that since we are only considering the case of zero orbital angular momentum, we do not observe any crossings of the energy levels, unlike in Ref.~\cite{PhysRevB_33_8336}.

\begin{figure}[t]
\begin{center}
\hspace{-1mm}
\includegraphics[trim = 3mm 1mm 4mm 2mm, clip, scale = 0.39]{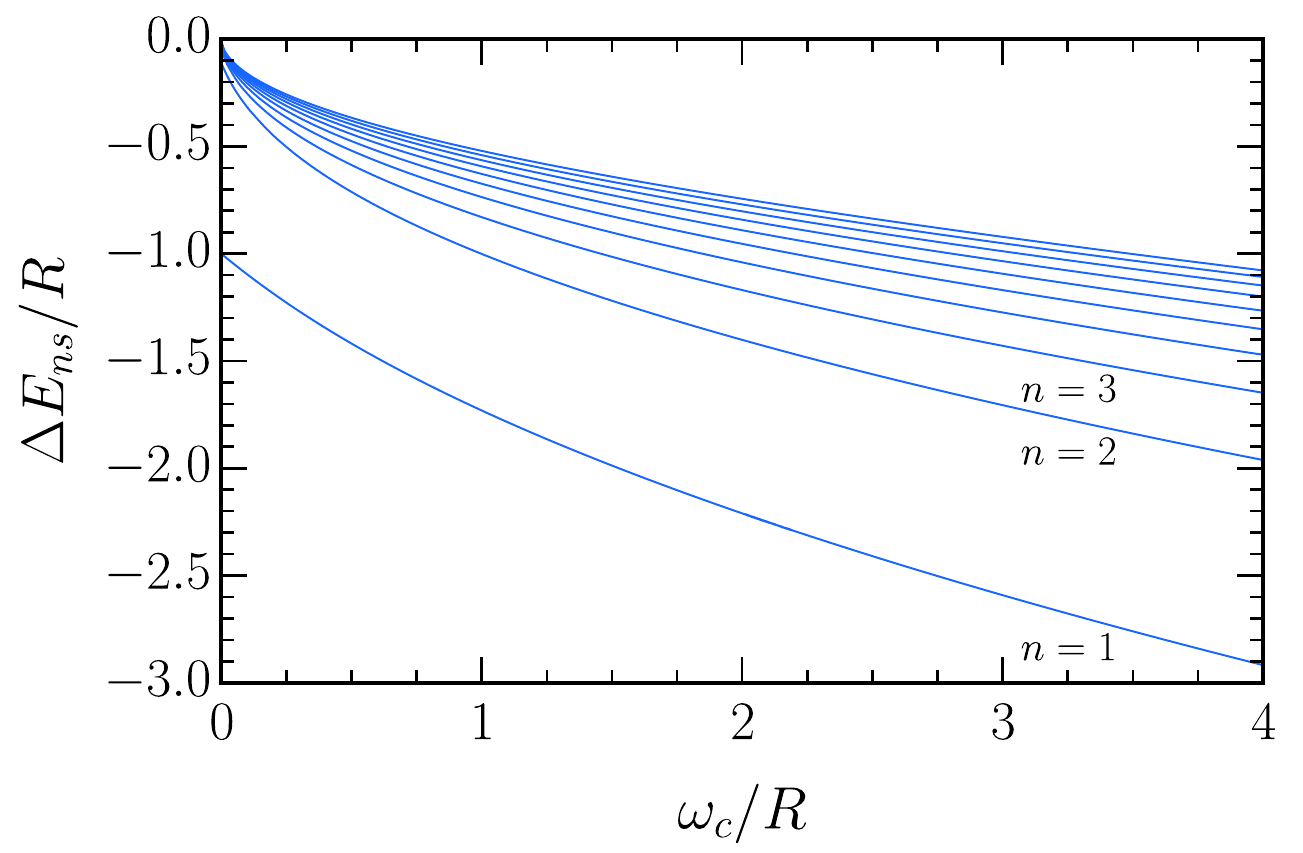}
\caption{Exciton energies from Fig.~\ref{fig:fig_exciton_2} measured with respect to the corresponding Landau energy levels of a free electron-hole pair, $\Delta{E}_{ns}\equiv{E}_{ns}-(2n-1)\hspace{0.05em}\omega_c$, where we label the states by the principal quantum number $n$.  This depiction reveals how the excitons become more tightly bound at higher fields.}
\label{fig:fig_exciton_1}
\end{center}
\end{figure}

High magnetic fields quench the kinetic energy of 2D electron systems so that the effect of the Coulomb interaction dominates --- see for instance Ref.~\cite{PhysB_177_401}.  To demonstrate this point, we plot the interaction-induced energy shifts in Fig.~\ref{fig:fig_exciton_1} by defining the exciton energies relative to the corresponding Landau energies of a free electron-hole pair.  This alternative representation of the solution to Eq.~\eqref{eq:exciton_kappa} is more intuitive, because it shows how the excitons become more strongly bound as the confinement due to the magnetic field increases.

\begin{figure}[t]
\begin{center}
\hspace{-2mm}
\includegraphics[trim = 28mm 2mm 26mm 21mm, clip, scale = 0.45]{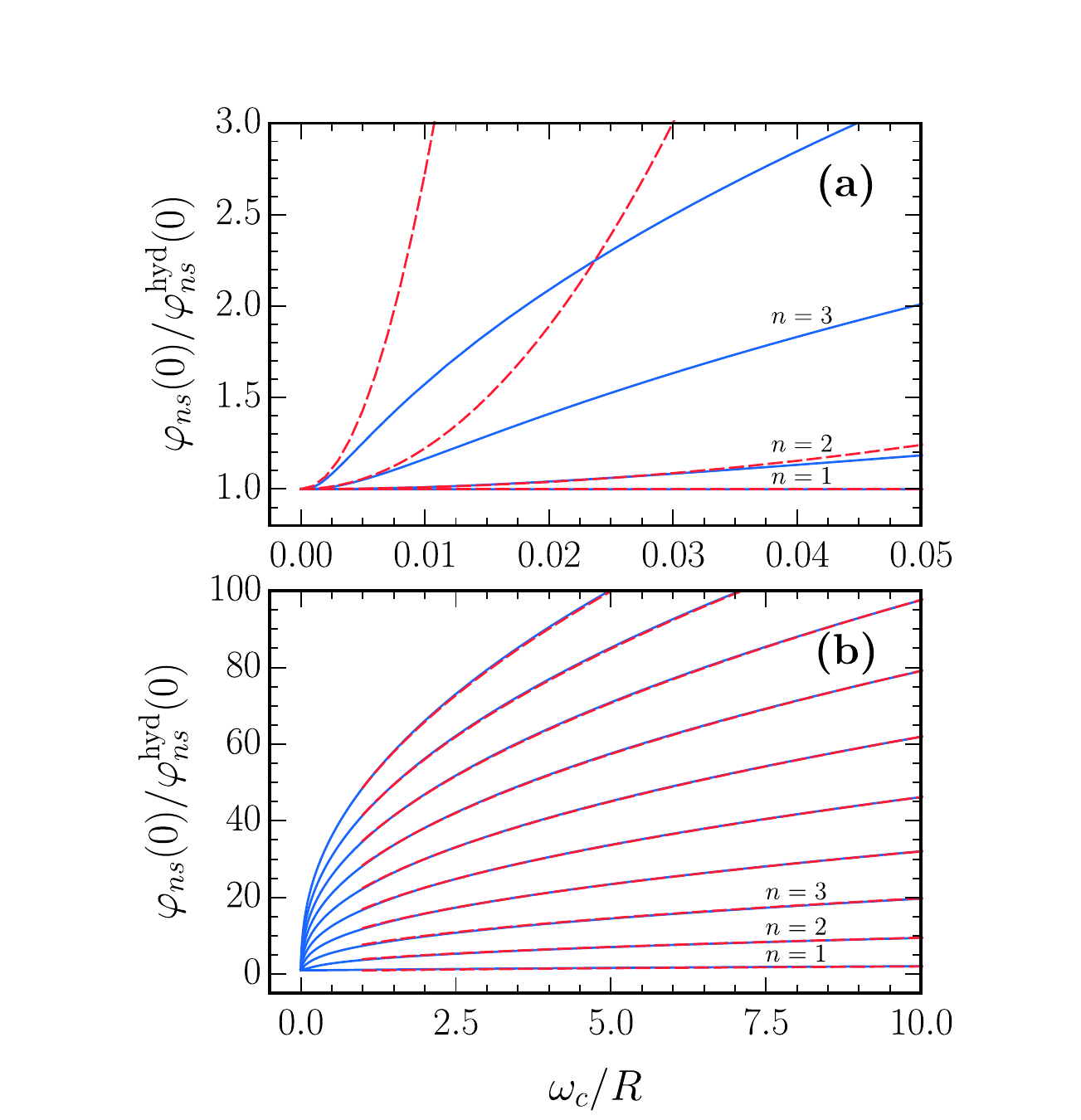}
\caption{Exciton wave function at zero separation as a function of magnetic field and normalized by its zero-field value.  For clarity, in the low-field limit (a) we show the first four lowest energy states, and in the high-field limit (b) the first\linebreak ten states.  The solid blue lines are calculated using Eq.~\eqref{eq:zerosep_WFs}, while the dashed red lines are given by the first-order perturbation theory results of Eqs.~\eqref{eq:low-B_WFs} and~\eqref{eq:high-B_WFs} in panels (a) and (b), respectively.}
\label{fig:fig_exciton_3}
\end{center}
\end{figure}

\subsection{Exciton wave function and oscillator strength}

\begin{figure*}[t]
\begin{center}
\hspace{-1mm}
\includegraphics[trim = 0mm 0mm 0mm 6mm, clip, scale = 0.36]{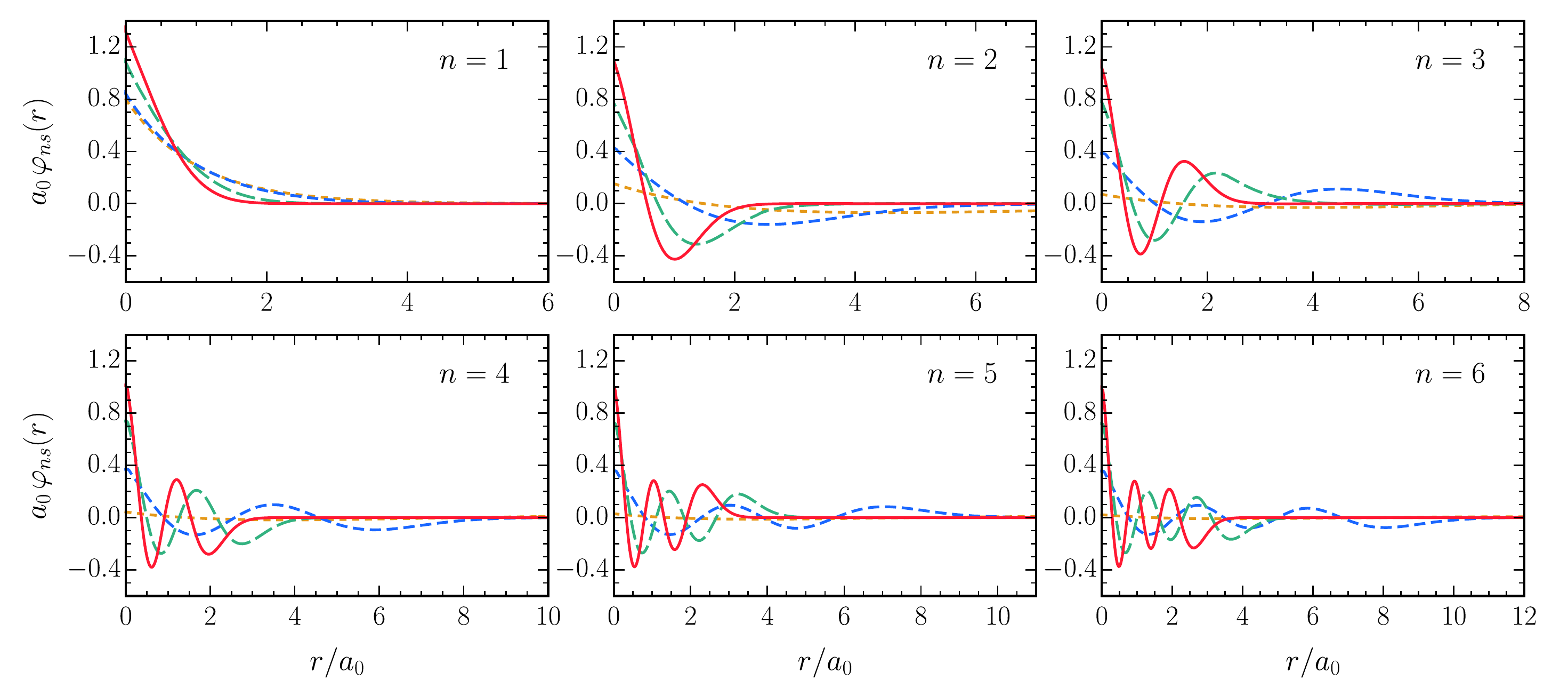}
\caption{Real-space wave functions for the $s$ series of Rydberg excitons as a function of electron-hole separation.  In each panel,\linebreak we show how the same state evolves with increasing external magnetic field:  $\omega_c/R=0.0$ [short-dashed orange], $0.5$ [medium-dashed blue], $2.5$ [long-dashed green], $5.0$ [solid red].  Note that the zero-field wave functions (in orange) for the $ns$ excitons also possess $n-1$ nodes or zero crossings.  However, these mostly occur beyond the range of the plots (at $r\gg10a_0$), and the wave function oscillations are also much smaller in magnitude, such that they are not visible.}
\label{fig:fig_exciton_5}
\end{center}
\end{figure*}

In addition to the energies, Eq.~\eqref{eq:exciton_kappa} allows us to extract the numerically exact electron-hole wave functions $\varphi(r)$ by finding the eigenfunctions $\widetilde\varphi_\kappa$.  The wave functions give access to the exciton oscillator strengths, which are proportional to the probabilities of the recombination of an electron and a hole $|\varphi(r=0)|^2$, and which thus determine the intensities of the exciton absorption peaks.  We should be careful to note that the obtained eigenfunctions need to be normalized according to the condition,
\begin{align}
\label{eq:ex_norm_cond}
1=\int{d}^2\r\,|\varphi(r)|^2=\int{d}^2\!\bm{\rho}\,\frac{4}{\rho}|\widetilde{\varphi}(\rho)|^2=8\pi\sum_{\bm\kappa,\,\bm\kappa'}\frac{\widetilde{\varphi}_{\kappa}\widetilde{\varphi}^*_{\kappa'}}{|\bm\kappa-\bm\kappa'|}\,.
\end{align}
To ensure this is the case, we take
\begin{align}
\label{eq:phinorm}
\widetilde\varphi_\kappa\to\widetilde\varphi_\kappa/{\mathcal{N}},
\end{align}
where ${\mathcal{N}}^2=8\pi\sum_{\bm\kappa,\,\bm\kappa'}\widetilde{\varphi}_{\kappa}\widetilde{\varphi}^*_{\kappa'}/|\bm\kappa-\bm\kappa'|$.  In the following, we implicitly assume that we are always working with the appropriately normalized solutions of Eq.~\eqref{eq:exciton_kappa}.

A key result is hence the magnetic-field dependence of the electron-hole wave function at zero separation in the original real space, $\varphi(r=0)$.  Since $r=0$ implies that $\rho=0$, this quantity can be simply calculated in $\bm\kappa$ space as shown below:
\begin{align}
\label{eq:zerosep_WFs}
\varphi(r=0)=\frac1{a_0}\widetilde{\varphi}(\rho=0)=\frac1{a_0}\sum_{\bm\kappa}\widetilde{\varphi}_{\kappa}\,.
\end{align}
For the remainder of this paper, we define all wave functions to be positive at $r=0$.

In Fig.~\ref{fig:fig_exciton_3}, we display the zero-separation electron-hole wave functions as a function of magnetic field.  On the vertical axis, we define $\varphi_{ns}(0)\equiv\varphi(r=0)$ where ``$ns$'' is the state label of the exciton, and we rescale this object by its value at zero field, which we know analytically from the wave functions of the 2D hydrogen problem $\varphi_{ns}^\mathrm{hyd}(0)$ --- see Eq.~\eqref{eq:hyd_WFs} below.  Akin to Fig.~\ref{fig:fig_exciton_2}, we compare our exact numerical results for the $ns$ exciton wave functions to the approximate perturbative results in the weak- (a) and strong-field (b) limits.  In both limits, we plot the sums of the zeroth- and first-order corrections to the appropriate unperturbed wave functions at $r=0$.

The relevant perturbative expression in the weak-field regime is given by
\begin{align}
\label{eq:low-B_WFs}
\varphi_{ns}(0)\simeq\hspace{+1.10mm}&\varphi_{ns}^{\mathrm{hyd}}(0)\hspace{+0.15mm}+\hspace{+0.20mm}\frac{\bar\omega_c^2}{4}\sum_{n'\hspace{+0.09mm}\neq\,n}\frac{1}{\bar{E}_{ns}^{\mathrm{hyd}}-\bar{E}_{n'\!s}^{\mathrm{hyd}}}\hspace{+0.55mm}\times\nn\\&\hspace{-1.35mm}\left[\frac{1}{a_0^2}\int{d}^2\r\,{r}^2\varphi_{ns}^{\mathrm{hyd}}(r)\varphi_{n'\!s}^{\mathrm{hyd}}(r)\right]\!\varphi_{n'\!s}^{\mathrm{hyd}}(0)\,,
\end{align}
which involves calculating the matrix elements of $r^2$ in the 2D $s$-wave hydrogenic basis.  Here, $\bar{E}_{ns}^{\mathrm{hyd}}$ are the 2D hydrogen energies and
\begin{align}
\label{eq:hyd_WFs}
\varphi_{ns}^{\mathrm{hyd}}(r)=\sqrt{\frac{2/\pi}{(2n-1)^3}}\,\mathrm{exp}\!\left(\hspace{-0.70mm}-\frac{r/a_0}{2n-1}\hspace{-0.70mm}\right)\!\mathrm{L}_{n-1}\!\left[\frac{2(r/a_0)}{2n-1}\right]\!\frac{1}{a_0}
\end{align}
are the corresponding wave functions, where $\mathrm{L}_n(x)$ represents the Laguerre polynomial, and $n$ again denotes the principal quantum number.

The equivalent perturbative expression in the strong-field regime is
\begin{align}
\label{eq:high-B_WFs}
\varphi_{N\hspace{-0.20mm}s}(0)\simeq\hspace{+0.15mm}\frac{1}{a_0}\sqrt{\frac{\bar\omega_c}{2\pi}}\hspace{+0.15mm}+\hspace{+0.20mm}\frac{1}{2\sqrt{2\pi}a_0}\sum_{N\hspace{-0.10mm}'\hspace{+0.10mm}\neq\,N}\frac{V_{N\hspace{-0.30mm},\hspace{+0.20mm}N\hspace{-0.10mm}'}^{s}}{N-N'}\,,
\end{align}
where $N$ is again the Landau level index.  We have arrived\linebreak at this expression  by using the 2D $s$-wave harmonic oscillator wave functions~\cite{PhysRevB_33_8336};  the first term is simply the value of these wave functions at the origin, while the second term involves calculating the matrix elements of $1/r$ in the harmonic oscillator basis~\cite{PhysRevB_33_8336}:
\begin{align}
\label{eq:Vfunc}
V_{N\hspace{-0.30mm},\hspace{+0.20mm}N\hspace{-0.10mm}'}^{s}=\hspace{+0.50mm}&-\sqrt{\frac{\pi}{2}}\,\hspace{+0.20mm}\mathrm{sec}(N\pi)\,\frac{\Gamma(\frac12-N+N')}{\Gamma(1+N)}\nn\\&\hspace{-0.25mm}\times\pFq[4]{3}{2}{\frac12,-N,\frac12-N+N'\hspace{-0.50mm}}{\frac12-N,1-N+N'\hspace{-0.50mm}}{1}\hspace{+0.10mm},
\end{align}
where ${_{p}\mathrm{F}_{\hspace{-0.25mm}q}}[\{a_1,\,...,\,a_q\};\{b_1,\,...,\,b_q\};x]$ represents the regularized generalized hypergeometric function.

In the upper panel (a) of Fig.~\ref{fig:fig_exciton_3}, we can see that perturbation theory is accurate for a decreasing range of magnetic fields as the principal quantum number $n$ increases --- the same trend as observed for the low-field energies in Fig.~\ref{fig:fig_exciton_2}(a).  In the lower panel (b), for clarity, we only show the perturbative results for $\bar\omega_c\geq1$ where they work well for all states considered.  For $\bar\omega_c<1$, we\linebreak have found that perturbation theory is accurate for an increasing range of magnetic fields as the Landau-level index $N$ increases --- the same trend as seen for the high-field energies in Fig.~\ref{fig:fig_exciton_2}(b).  However, the wave functions at moderate field strengths, which is often the relevant regime for experiments, cannot be described by perturbative methods up to first order.

In fact, we can extract the entire relative electron-hole wave functions in the original real space by carrying out the reverse series of transformations $\kappa\to\rho\to{r}$ on the eigenfunctions $\widetilde{\varphi}_{\kappa}$ of Eq.~\eqref{eq:exciton_kappa}.  After normalization, we numerically perform the Fourier transform back into the rescaled real space, $\widetilde\varphi(\rho)=\int\!{d}^{\hspace{0.05em}2}\!\hspace{0.10em}\bm\kappa/(2\pi)^2\,{e}^{i\bm\kappa\cdot\bm\rho}\,\widetilde{\varphi}_{\kappa}$, which then enables us to obtain $\varphi(r)=\widetilde\varphi(\sqrt{8a_0^2\rho}\,)$.  We depict these wave functions $\varphi(r)$ for a range of excitonic states and magnetic fields in Fig.~\ref{fig:fig_exciton_5}.  Upon applying a magnetic field of increasing strength, we observe that the oscillations of the wave functions increase in magnitude and also shift closer to the origin.  (Note, the zero-field lines in each panel possess the same number of oscillations as the others, but they are much smaller in magnitude and extend far beyond the range of the plots.)  In particular, the large upshifts of the exciton oscillator strengths $\varphi_{ns}(0)$ --- which are also visible in Fig.~\ref{fig:fig_exciton_3} --- correspond to stronger light-matter coupling in the Rydberg polariton systems of Sec.~\ref{sec:Exciton-Polariton_Problem}.  At higher fields, we furthermore see that the wave functions decay more rapidly to zero at longer separations; these exponential suppressions being due to their harmonic oscillator nature in this limit.

\begin{figure}[t]
\begin{center}
\hspace{-0.5mm}
\includegraphics[trim = 0mm 1mm 0mm 2mm, clip, scale = 0.39]{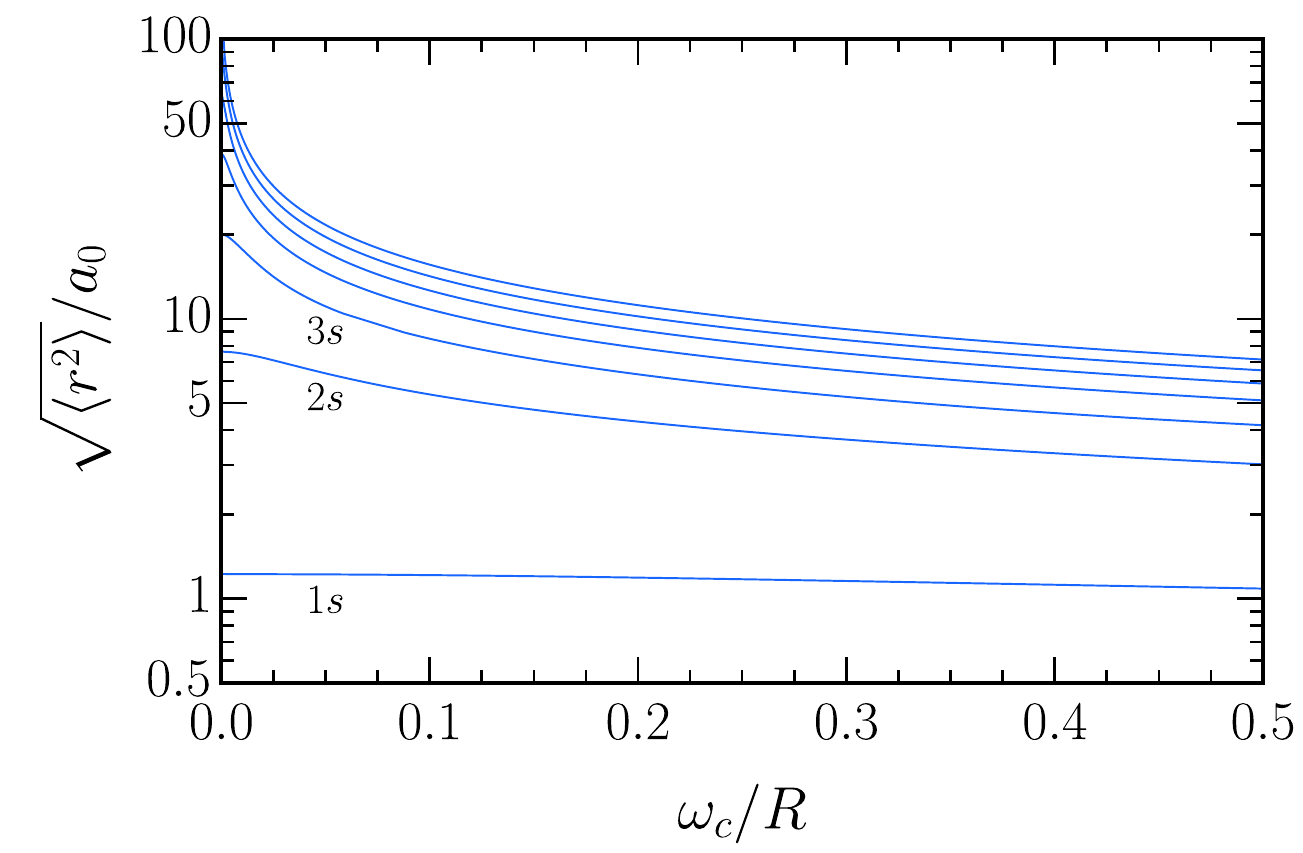}
\caption{Root-mean-square electron-hole separation as a fun-\linebreak ction of magnetic field for the seven lowest energy $s$ excitons.}
\label{fig:fig_exciton_4}
\end{center}
\end{figure}

To further quantify how the wave functions of the Rydberg states shrink with the applied magnetic field, we determine the average squared electron-hole separation. This expectation value straightforwardly transforms into $\bm\kappa$ space and can thus be examined within our approach:
\begin{align}
\label{eq:r_squared_X}
\frac{\left<r^2\right>}{a_0^2}\equiv\frac{1}{a_0^2}\int d^2\r\, r^2|\varphi(r)|^2
=
32\sum_{\bm\kappa}|\widetilde{\varphi}_{\kappa}|^2\,.
\end{align}
The result is presented in Fig.~\ref{fig:fig_exciton_4}, which again shows how the excited states are strongly modified even at relatively small ratios of $\omega_c/R$.

\section{Exciton-Polariton in a Magnetic Field}
\label{sec:Exciton-Polariton_Problem}

We now turn to the case of a strong light-matter coupling, which can be realized by embedding a 2D semiconductor in an optical microcavity~\cite{Microcavities}.  Cavity photons are localized between Bragg reflectors, and are cyclically absorbed and emitted by the excitons in the semiconductor.  Once the rate of energy exchange between photons and excitons exceeds the loss rate, the system becomes characterized by new eigenmodes, i.e., polaritons, which are quantum mechanical superpositions of light and matter \cite{Keeling_2007,RevModPhys.82.1489,RevModPhys.85.299}.  In this section, we thus evoke the full Hamiltonian~\eqref{eq:Hamiltonian} of Sec.~\ref{sec:Model} and we exactly solve the problem of a two-dimensional exciton-polariton in a static transverse magnetic field.  In the literature, a single polariton has only been (numerically) exactly treated in 2D momentum space and in the absence of a magnetic field~\cite{PhysRevRes_1_033120}.  Below, we show how to connect that solution with the transformations of the preceding section.  As before, our method directly applies to thin quantum-well semiconductors, and can be adapted to describe atomically thin materials by making only minor adjustments.

We begin by writing down the most general state for a single exciton-polariton at zero center-of-mass momentum,
\begin{align}
\label{eq:pol_WF}
\ket\psi=\int{d}^2\r\,\varphi(r)\ket\r\otimes\ket0+\gamma\ket{\mathrm{vac}}\otimes\ket1\,.
\end{align}
This imposes the rotating wave approximation, which is reasonable since the bandgap is much larger than the energy scale of the light-matter coupling.  Above, the basis\linebreak ket $\ket\r$ corresponds to an electron-hole pair at $\mathbf{K}=0$ and relative separation $\r$, $\ket{\mathrm{vac}}$ denotes the electron-hole vacuum, $\ket0$ is the photon vacuum, and $\ket1=\hat{c}^\dagger\!\ket0$.  In addition, $\varphi(r)$ is the $s$-wave real-space wave function for the relative motion of the electron and hole, while $\gamma$ is the photon amplitude, and we assume that the polariton state is normalized such that
\begin{align}
\label{eq:pol_NC}
1=\langle\hspace{+0.04em}\psi\hspace{+0.06em}|\hspace{+0.05em}\psi\hspace{+0.05em}\rangle=\int{d}^2\r\,|\varphi(r)|^2+|\gamma|^2\,.
\end{align}

By projecting the Schr\"{o}dinger equation onto the subspaces of Eq.~\eqref{eq:pol_WF}, viz.,
\begin{subequations}
\begin{align}
\label{eq:proj_SE}
\bra\r\otimes\bra0\!\hspace{0.10mm}\left(E-\hat{H}_\mathrm{eff}'\hspace{0.15mm}\right)\!\hspace{0.10mm}\ket\psi&=0\,,\\\bra{\mathrm{vac}}\otimes\bra1\!\hspace{0.10mm}\left(E-\hat{H}_\mathrm{eff}'\hspace{0.15mm}\right)\!\hspace{0.10mm}\ket\psi&=0\,,
\end{align}
\end{subequations}
we obtain two coupled eigenvalue equations for the energy $E$ of the polariton:
\begin{subequations}
\label{eq:phirH_eqs}
\begin{align}
\label{eq:phirH_eq1}
&\hspace{-1.50mm}\left[E+\frac1{2\mu}\!\left(\frac{d^2}{dr^2}+\frac1r\frac{d}{dr}\right)\!-\frac{\mu\omega_c^2}2r^2+\frac{e^2}\varepsilon\frac1r\right]\!\varphi(r)=g\gamma\hspace{+0.10em}\delta^2(\r)\,,\\
\label{eq:phirH_eq2}
&\hspace{-0.55mm}\big(E-\omega\big)\gamma=g\int{d}^2\r\,\varphi(r)\hspace{+0.10em}\delta^2(\r)\,.
\end{align}
\end{subequations}
Similar to the previous section, we measure all energies (including the bare photon frequency $\omega$) with respect to the semiconductor bandgap.

The presence of the Dirac delta function in Eq.~\eqref{eq:phirH_eq1} forces the electron-hole wave function $\varphi(r)$ to diverge logarithmically at short range.  It is convenient to explicitly separate out the divergent part as follows:
\begin{align}
\label{eq:varphi(r)}
\varphi(r)=\beta(r)-\frac{g\gamma\mu}{\pi}\hspace{+0.1em}\mathrm{K}_0(r/a_0)\,,
\end{align}
where $\beta(r)$ denotes a regular function and $\mathrm{K}_0(r/a_0)$ is the zeroth-order modified Bessel function of the second kind. This takes advantage of how $\mathrm{K}_0$ is the Green's function of free-particle motion in two dimensions.  In other words, it satisfies $\big[E_{1s}^\text{hyd}+\nabla^2_\mathbf{r}/(2\mu)\big]\mathrm{K}_0\big(r/a_0\big)=-\pi\delta^2(\r)/\mu\hspace{+0.05em}$, and thus it explicitly cancels the delta function in Eq.~\eqref{eq:phirH_eq1}. However, $\mathrm{K}_0(r/a_0)$ diverges as $\ln(r/a_0)$ at $r\to0$, which means that the integral in Eq.~\eqref{eq:phirH_eq2} is formally divergent.  We require a renormalization procedure to remove this infinity~\cite{Levinsen2015review}, and fortunately, such a procedure has already been elucidated for the case of zero magnetic field in Ref.~\cite{PhysRevRes_1_033120}.  Because the divergence here occurs at zero electron-hole separation where the magnetic field has no effect, we can renormalize our system of equations in precisely the same manner as was done in that paper.

Experimentally, the exciton-polariton system is typically characterized at zero magnetic field by fitting the observed spectrum to that of a system of coupled oscillators~\cite{RevModPhys.85.299}. For the 1$s$ exciton at energy $E_{1s}^\text{hyd}$ and the phot-\linebreak on at energy $E_{1s}^\text{hyd}+\delta$, coupled by the light-matter (Rabi) coupling strength $\Omega$, the expected spectrum is
\begin{align}\label{eq:two-level dispersion}
E_{{\tiny  \begin{matrix} \text{LP}\\\text{UP}\end{matrix}}} &=E_{1s}^\text{hyd}+\frac12\!\left(\delta\mp\sqrt{\delta^2+4\Omega^2}\,\right),
\end{align}
where LP and UP correspond to the lower and upper polaritons, respectively.  Due to the form of the wave function in Eq.~\eqref{eq:varphi(r)}, it was explained in Ref.~\cite{PhysRevRes_1_033120} that the exciton-photon detuning $\delta$ is not simply related to $\omega$ and $E_{1s}^\text{hyd}=-R$. Instead, within our theory we obtain a re-\linebreak normalized detuning between the $1s$ exciton and the cavity photon energies~\cite{PhysRevRes_1_033120}:
\begin{align}
\label{eq:detuning}
\delta=\omega-\frac{g^2\mu}{\pi}\int{d}^2\r\,\mathrm{K_0}(r/a_0)\hspace{+0.10em}\delta^2(\r)+R\,.
\end{align}
Although the integral above is formally divergent, it exactly cancels with the divergence on the right-hand side of Eq.~\eqref{eq:phirH_eq2}, when that equation is written in terms of $\delta$ rather than $\omega$.  It is important to emphasize that Eq.~\eqref{eq:detuning} corresponds to the actual physical exciton-photon detuning that one would observe in experiment.  Similarly, in order for Eq.~\eqref{eq:phirH_eqs} to recover the LP and UP energies of the coupled oscillator model \eqref{eq:two-level dispersion} at zero magnetic field, the LP-UP Rabi coupling can be defined in terms of the contact coefficient $g$ via
\begin{align}
\label{eq:Rabi}
    \Omega=g\hspace{0.05em}\varphi^\mathrm{hyd}_{1s}(0)=\frac g{a_0}\sqrt{\frac2\pi}\,.
\end{align}
The expressions \eqref{eq:detuning} and \eqref{eq:Rabi} that relate the experimental observables $\delta$ and $\Omega$ to the parameters of the model have been shown to work very well when $|\delta|,\,\Omega\lesssim{R}$~\cite{PhysRevRes_1_033120}.  In the following, we use these observables to characterize the polariton system.

\begin{figure*}[t]
\begin{center}
\hspace{-1mm}
\includegraphics[trim = 0mm 2mm 0mm 1mm, clip, scale = 0.36]{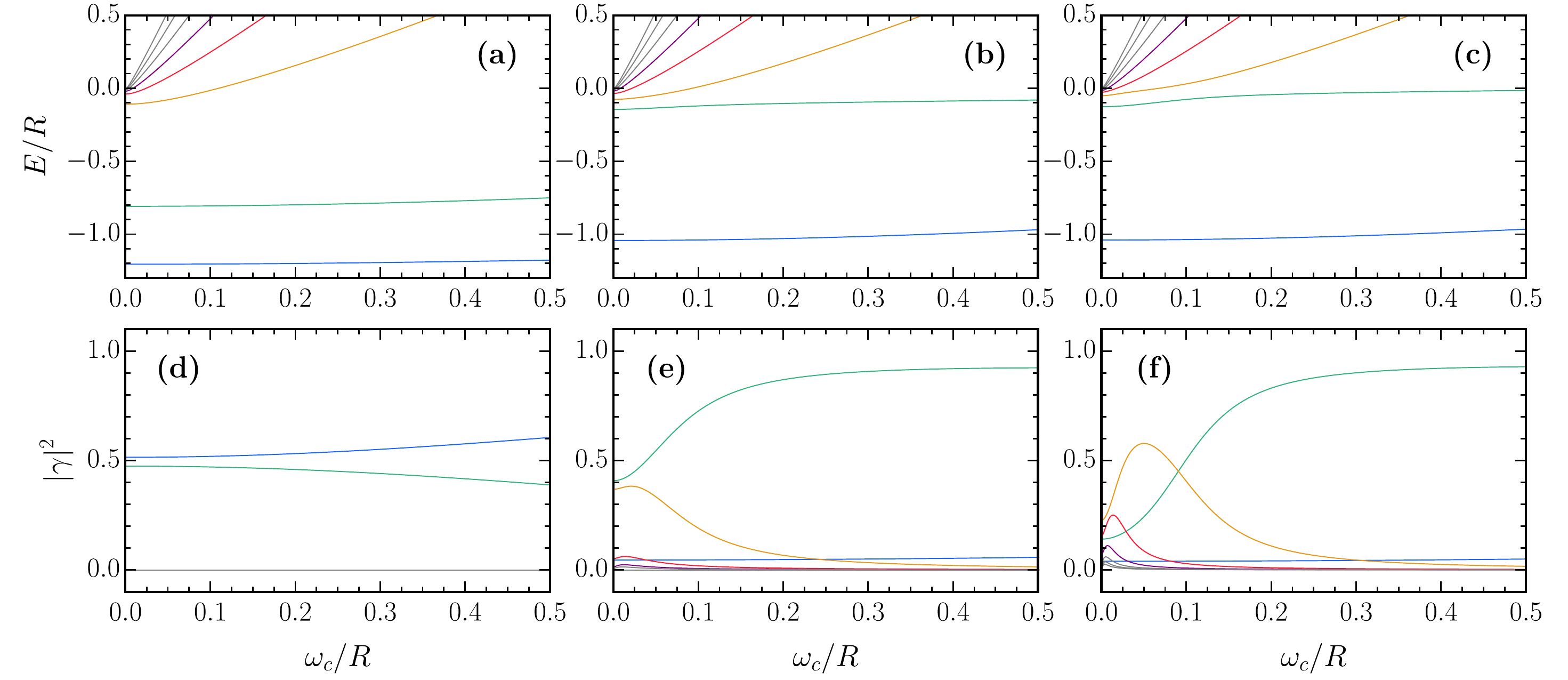}
\caption{Upper panels:  polariton energy as a function of magnetic field for the eight lowest energy $s$ states.  The Rabi coupling strength at zero field is $\Omega/R=0.20$ in all cases (a)--(c), while the exciton-photon detuning at zero field increases from left to right:  (a) $\delta/R=0$, (b) $\delta/R=8/9 =\bar{E}_{2s}^{\text{hyd}}-\bar{E}_{1s}^{\text{hyd}}$\hspace{-0.1em}, (c) $\delta/R=24/25 =\bar{E}_{3s}^{\text{hyd}}-\bar{E}_{1s}^{\text{hyd}}$\hspace{-0.1em}.  Lower panels:  corresponding photon frac-\linebreak tions, where the states have the same respective colors as in the upper panels.  In panel (d), note that because the photon is predominantly coupled to the $1s$ exciton, the photon fractions of the lower (blue) and upper (green) polaritons sum almost to one, while the photon fractions for the remaining states are essentially zero on this scale.}
\label{fig:fig_pol_new}
\end{center}
\end{figure*}

To proceed, we carry out the replacements $\omega\to\delta$ and $g\to\Omega$, and then we perform the same variable transformation as in the exciton problem, $r\to\rho=r^2/(8a_0^2)$.  Eventually, we arrive at the dimensionless coupled equations shown below:
\begin{subequations}
\label{eq:phirho_eqs}
\begin{align}
\label{eq:phirho_eq1}
&\hspace{-1.50mm}\left[\frac{2\hspace{0.01em}\bar{E}}{\rho}\hspace{-0.1mm}+\hspace{-0.1mm}\frac{d^2}{d\rho^2}\hspace{-0.1mm}+\hspace{-0.1mm}\frac1\rho\frac{d}{d\rho}\hspace{-0.1mm}-\hspace{-0.1mm}4\hspace{0.10em}\bar{\omega}_c^2\hspace{-0.1mm}+\hspace{-0.1mm}\frac{1}{\sqrt{2\hspace{0.01em}\rho^3}}\right]\!\widetilde{\varphi}(\rho)\hspace{-0.2mm}=\hspace{-0.2mm}\sqrt{\frac\pi8}\hspace{0.10em}\bar\Omega\hspace{0.10em}\gamma\hspace{0.10em}\delta^2(\bm\rho)\,,\\
\label{eq:phirho_eq2}
&\hspace{-1.40mm}\left[\bar{E}-\bar{\delta}-\frac{\bar\Omega^2}{4}\int{d}^2\!\bm\rho\,\mathrm{K_0}\!\left(2\sqrt{2\rho}\hspace{0.10em}\right)\!\hspace{0.05em}\delta^2(\bm\rho)+1\right]\!\gamma\nn\\&\hspace{-1.15mm}=\sqrt{\frac\pi2}\hspace{0.10em}\bar\Omega\int{d}^2\!\bm\rho\,\widetilde{\varphi}(\rho)\hspace{0.10em}\delta^2(\bm\rho)\,,
\end{align}
\end{subequations}
with $\bar\Omega\equiv\Omega/R$ and $\bar{\delta}\equiv\delta/R$~\footnote{In addition to the change of variable, we rescale the upper equation~\eqref{eq:phirho_eq1} by $2a_0/(R\rho)$.}.  Notably, this system of equations is now fully described in terms of the experimentally measurable parameters of the light-matter coupled system, $\Omega$ and $\delta$, at a finite magnetic field with cyclotron frequency $\omega_c$.

Subsequently, we Fourier transform Eq.~\eqref{eq:phirho_eqs} into the corresponding dimensionless momentum $\bm\kappa$ space, yielding
\begin{subequations}
\label{eq:phikappa_eqs_final}
\begin{align}
\label{eq:phikappa_eq1_final}
&\hspace{-0.1mm}\bar{E}\sum_{\bm\kappa'}\frac{4\pi\hspace{0.1em}\widetilde{\varphi}_{\kappa'}}{|\bm\kappa-\bm\kappa'|}-\kappa^2\hspace{0.1em}\widetilde{\varphi}_{\kappa}-4\hspace{0.1em}\bar{\omega}_c^2\hspace{0.1em}\widetilde{\varphi}_{\kappa}+\frac{\Gamma(1/4)}{\Gamma(3/4)}\sum_{\bm\kappa'}\frac{\pi\hspace{0.1em}\widetilde{\varphi}_{\kappa'}}{\sqrt{|\bm\kappa-\bm\kappa'|}}\nn\\&\hspace{-1.1mm}=\sqrt{\frac\pi8}\hspace{0.1em}\bar\Omega\hspace{0.1em}\gamma\,,\\
\label{eq:phikappa_eq2_final}
&\hspace{-1.5mm}\left[\bar{E}-\bar{\delta}-\frac{\pi\hspace{0.1em}\bar\Omega^2}{4}\hspace{-0.3mm}\sum_{\bm\kappa}\hspace{-0.3mm}\left\{\hspace{-0.2mm}\frac{1}{\kappa^2}+\frac{\pi}{\kappa^3}\hspace{-0.2mm}\!\left[\hspace{0.1em}\hspace{-0.4mm}\mathrm{Y}_0\!\left(\frac{2}{\kappa}\right)\!-\mathrm{H}_0\!\left(\frac{2}{\kappa}\right)\hspace{-0.3mm}\right]\hspace{-0.5mm}\right\}\hspace{-0.2mm}+1\right]\!\gamma\nn\\&\hspace{-1.1mm}=\sqrt{\frac\pi2}\hspace{0.1em}\bar\Omega\sum_{\bm\kappa}\widetilde{\varphi}_{\kappa}\,,
\end{align}
\end{subequations}
where $\mathrm{Y}_0(2/\kappa)$ is the zeroth-order Bessel function of the second kind and $\mathrm{H}_0(2/\kappa)$ is the zeroth-order Struve function. As with Eq.~\eqref{eq:exciton_kappa}, Eq.~\eqref{eq:phikappa_eqs_final} represents a key result of this work, since it allows a numerically exact solution of the single-polariton problem in a magnetic field.

To solve this set of equations~\eqref{eq:phikappa_eqs_final} in $\bm\kappa$ space, we re-arrange Eq.~\eqref{eq:phikappa_eq2_final} for the photon amplitude $\gamma$ and substitute the result into Eq.~\eqref{eq:phikappa_eq1_final}.  Then we have a single eigenvalue-matrix equation where $\widetilde{\varphi}_{\kappa}$ is the eigenvector.  It is important to note that all sums on $\kappa$ need to be evaluated on the same quadrature grid so that the divergences are properly cancelled.  To plot the polariton spectrum at fixed light-matter coupling strength $\Omega$ and detuning $\delta$, for instance, we can input a range of energies $E$, and for each one obtain the (squared) cyclotron frequency $\omega_c$ as the eigenvalue.  Similar to the previous section, here we have considered the 2D Coulomb interactions characteristic of GaAs quantum wells.  However, the consideration of other interaction potentials is also possible and this would only affect the rightmost term on the left-hand side of Eq.~\eqref{eq:phikappa_eq1_final} --- refer to Appendix~\ref{sec:Appendix_B}.

As in the exciton problem, we must normalize our solution. This is now more complicated, since we have both the electron-hole wave function and the photon amplitude, which need to be simultaneously normalized according to Eq.~\eqref{eq:pol_NC}. A suitable approach is to first calculate the (unnormalized) photon amplitude from our (unnormalized) $\widetilde\varphi_\kappa$ according to Eq.~\eqref{eq:phikappa_eq2_final}:
\begin{multline}
\label{eq:extra_ch7}
\gamma=\Bigg(\bar{E}-\bar{\delta}-\frac{\pi\hspace{0.1em}\bar{\Omega}^2}{4}\sum_{\bm\kappa}\bigg\{\frac{1}{\kappa^2}\hspace{+0.80mm}+\\\frac{\pi}{\kappa^3}\!\left[\hspace{0.1em}\mathrm{Y}_0\!\left(\frac{2}{\kappa}\right)\!-\mathrm{H}_0\!\left(\frac{2}{\kappa}\right)\right]\!\bigg\}+1\Bigg)^{\hspace{-0.25mm}\!\!\!-1}\!\sqrt{\frac\pi2}\hspace{0.1em}\bar\Omega\sum_{\bm\kappa}\widetilde{\varphi}_{\kappa}\,.
\end{multline}
Similar to in Sec.~\ref{sec:Exciton_Problem}, we then rescale our numerical solution as follows: 
\begin{align}
    \widetilde\varphi_\kappa\to\widetilde\varphi_\kappa/{\mathcal N},\qquad \gamma\to\gamma/{\mathcal N},
\end{align}
with normalization
${\mathcal N}^2=|\gamma|^2+8\pi\sum_{\bm\kappa,\,\bm\kappa'}\widetilde{\varphi}_{\kappa}\widetilde{\varphi}^*_{\kappa'}/|\bm\kappa-\bm\kappa'|$ [compare with Eqs.~\eqref{eq:ex_norm_cond} and~\eqref{eq:phinorm}].  For the remainder of this paper, we implicitly assume this normalization for our solutions.

\begin{figure}[b]
\begin{center}
\hspace{-1mm}
\includegraphics[trim = 25mm 2mm 26mm 21mm, clip, scale = 0.45]{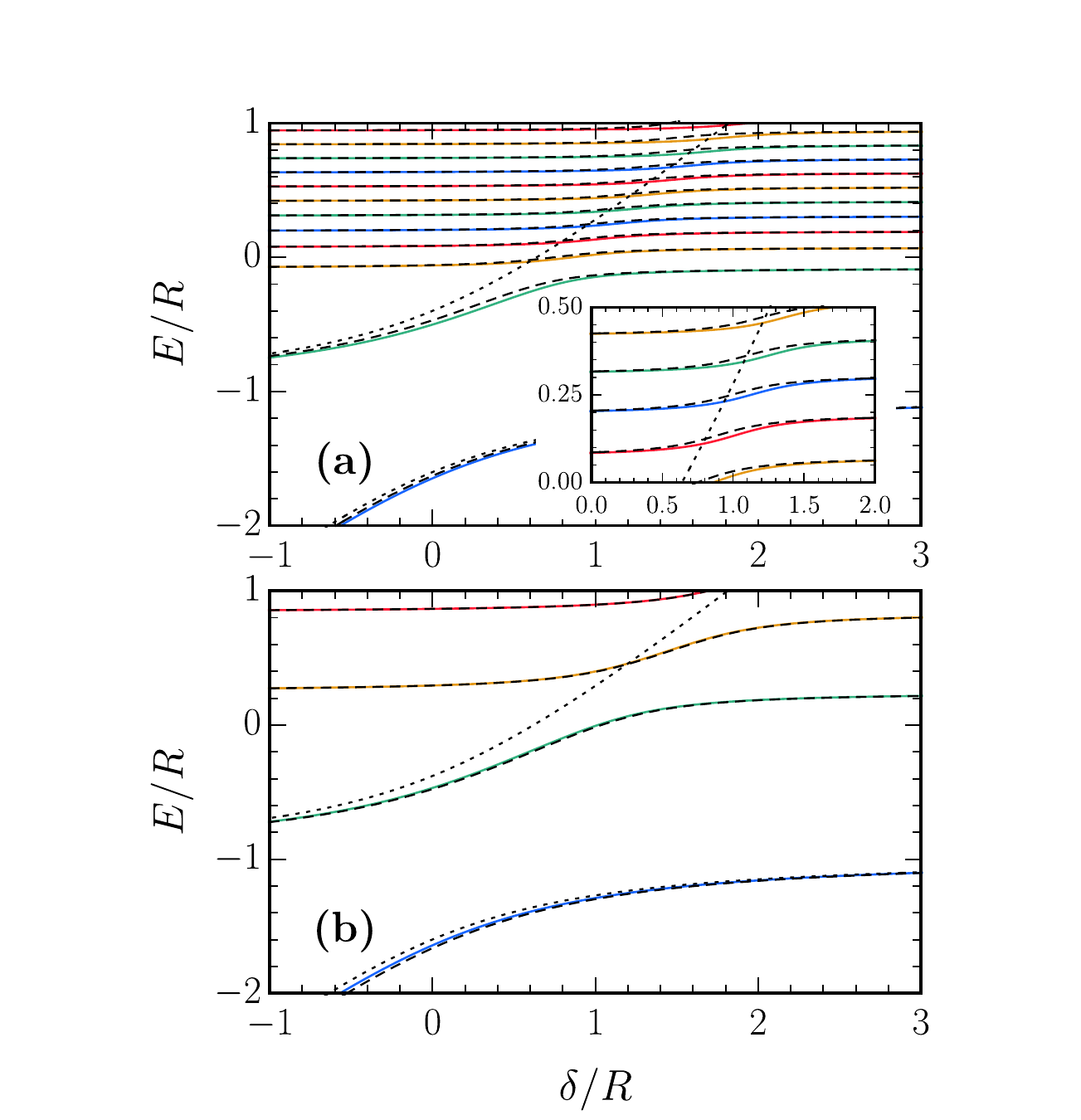}
\caption{Polariton energies in an external magnetic field as a function of exciton-photon detuning.  We compare our exact numerics, shown as solid colored lines, against the results of multilevel coupled oscillator models~\eqref{eq:COM} --- shown as dotted black lines for a $2$-level COM, and dashed black lines for a $12$-level COM.  The magnetic field strength is $\omega_c/R=0.05$ in (a) and $\omega_c/R=0.25$ in (b), while the Rabi coupling strength is $\Omega/R=0.60$ in both panels.}
\label{fig:fig_pol_COMs}
\end{center}
\end{figure}

\subsection{Results}

Figures~\ref{fig:fig_pol_new} and~\ref{fig:fig_pol_COMs} display our numerically exact spectra of exciton-polaritons in the presence of a magnetic field.  First, the upper and lower panels of Fig.~\ref{fig:fig_pol_new} show how the energies and photon fractions of the polariton branches evolve with the cyclotron frequency, $\omega_c/R$, when the resonant cavity frequency is tuned to the $(B=0)$ $1s$, $2s$, or $3s$ exciton states.  The upper panels (a)--(c) can therefore be directly compared against the corresponding spectrum in the absence of light-matter coupling, Fig.~\ref{fig:fig_exciton_2}(b), where the resonant cavity frequency would correspond to a horizontal line since it is independent of magnetic field.  In Fig.~\ref{fig:fig_pol_new}, we have used a light-matter coupling strength of $\Omega/R=0.2$, which is a typical value in current microcavity heterostructures.  The light-matter coupling leads to a splitting of energies close to the photon energy.  When the detuning is chosen so that the photon energy is resonant with the $1s$ state, we see in panel (d) that there is little coupling to any other excitonic states --- i.e., the photon fractions of the lower (blue) and upper (green) polaritons sum almost to one, while the photon fractions for the remaining states are essentially zero.  However, because the higher Rydberg energies are more closely spaced, when the photon is resonant with the $2s$ and $3s$ excitons, we see from the corresponding photon fractions in panels (e) and (f) that we begin to strongly couple to multiple excitonic states, and that these states can be tuned in and out of resonance via the magnetic field.  Interestingly, this coupling between excited states leads to a non-monotonic behavior of the photon fraction as a function of magnetic field.

Alternatively, we can consider the polariton spectra at fixed magnetic field as a function of exciton-photon detuning, as shown in Fig.~\ref{fig:fig_pol_COMs}.  We can infer the importance of strong light-matter coupling effects on the $ns$ exciton state by comparing our exact results to an $(n+1)$-level coupled oscillator model (COM).  In the COM, the exciton is assumed to be unperturbed by the strong coupling to light, and the polariton energies are obtained by diagonalizing the $(n+1) \times (n+1)$ matrix
\begin{equation}
    \begin{pmatrix}
        \delta + E_{1s}^{\text{hyd}} & \Omega_{1s} & \Omega_{2s} & \cdots & \Omega_{ns}\\ 
        \Omega_{1s} &
        E_{1s} & 0 & \cdots & 0 \\
        \Omega_{2s} & 0 & E_{2s} & \cdots & 0 \\ \vdots
         & \vdots & \vdots & \ddots & \vdots \\
         \Omega_{ns} & 0 & 0 & \cdots & E_{ns}
    \end{pmatrix},
\label{eq:COM}
\end{equation}
where $\Omega_{ns}=\Omega\hspace{0.05em}\varphi_{ns} (0)/\varphi_{1s}^{\text{hyd}} (0)$.  In Fig.~\ref{fig:fig_pol_COMs}, we can observe that for the moderate magnetic field case in panel (a), a 12-level coupled oscillator model deviates from the exact results in the regime of strong coupling between light and matter, with deviations of order $\sim\!15\%$ of $R$ for the chosen values of $\Omega$ and $\omega_c$, particularly evident in the inset. Upon increasing the magnetic field from (a) to (b), the system becomes more perturbative in the light-matter coupling, i.e., better described by a COM.  This can be understood as due to the larger difference between consecutive exciton energies (Fig.~\ref{fig:fig_exciton_2}), relative to the exciton coupling strength to light (Fig.~\ref{fig:fig_exciton_3}).  In particular, in the regime of strong magnetic field, Eqs.~\eqref{eq:strong-field_energies} and~\eqref{eq:high-B_WFs} together imply that $\Omega_{ns}/\omega_c\sim\bar{\omega}_c^{-1/2}$.

Note that when it comes to examining the polariton diamagnetic shifts (i.e., the change in energy as a function of magnetic field), the deviations of the COM become much more substantial, leading to even qualitatively incorrect results at only moderate values of $\Omega/R$.  We discuss this point further in Sec.~\ref{sec:Theoretical_Comparison_to_Experiment}, where we compare our exact numerics to recent experimental measurements.

We can quantify changes to the electron-hole (matter) part of the polariton due to light and to a magnetic field by computing the average squared electron-hole separation.  Similar to Eq.~\eqref{eq:r_squared_X} for the exciton, we have
\begin{align}
\label{eq:r_squared}
\frac{\left<r^2\right>_{\hspace{-0.5mm}\varphi}}{a_0^2}\equiv\int{d}^2\r\,\frac{r^2\hspace{+0.1em}|\varphi(r)|^2}{a_0^2\hspace{+0.1em}(1-|\gamma|^2)}=\sum_{\bm\kappa}\frac{32\hspace{+0.1em}|\widetilde{\varphi}_{\kappa}|^2}{1-|\gamma|^2}\,.
\end{align}
Importantly, to clearly show how the electron-hole wave function changes, we take the average within the matter-only part of the polariton wave function (indicated here by the subscript $\varphi$), and the division by the exciton fraction $1-|\gamma|^2=\int{d}^2\r\,|\varphi(r)|^2$ ensures that this matter part is normalized to unity.

\begin{figure*}[t]
\begin{center}
\hspace{-1mm}
\includegraphics[trim = 0mm 10mm 0mm 10mm, clip, scale = 0.36]{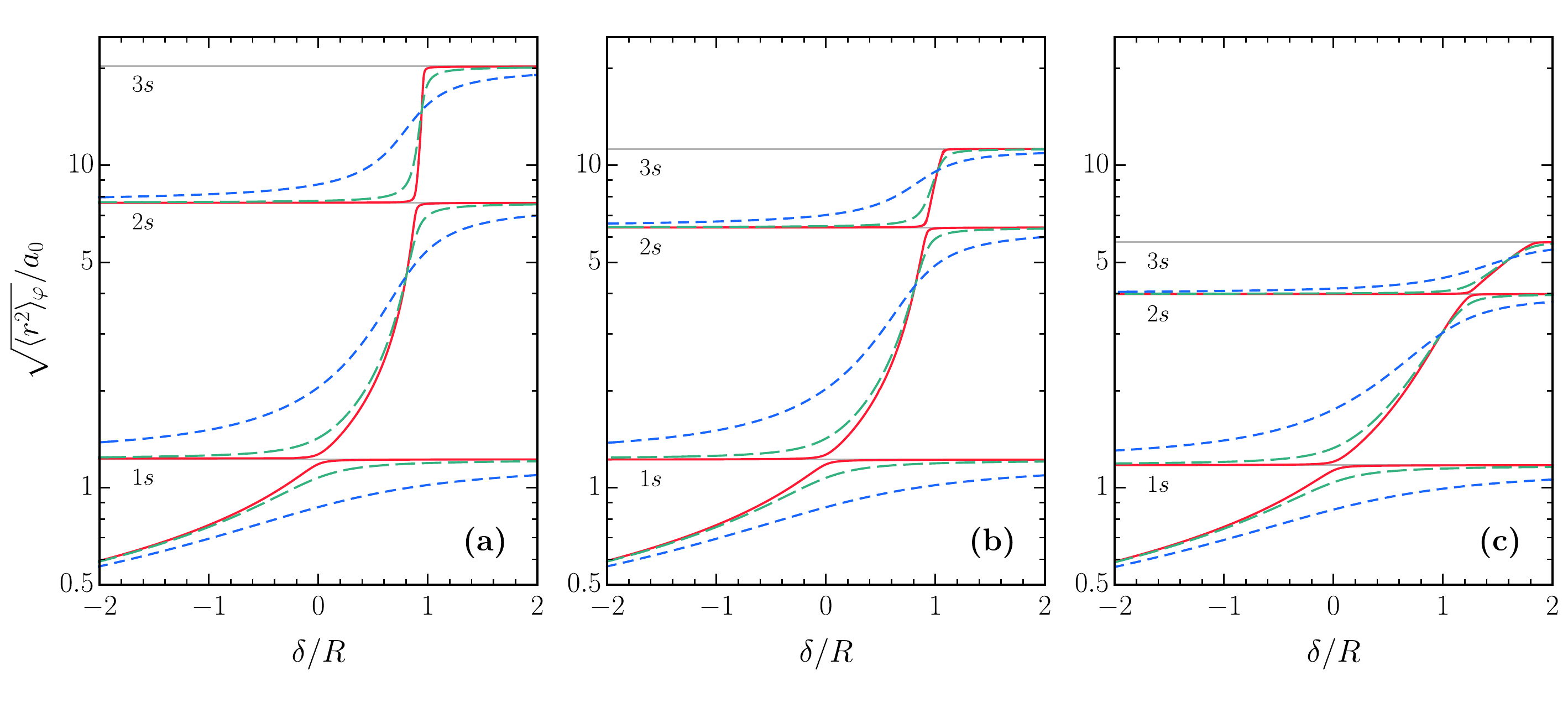}
\caption{Root-mean-square  electron-hole separation as a function of exciton-photon detuning for the three lowest energy polar-\linebreak iton states.  The magnetic field strength increases from left to right:  $\omega_c/R=0.00$ (a), $0.05$ (b), $0.25$ (c).  In all panels, the Rabi coupling strengths are $\Omega/R=0.05$ [solid red], $0.20$ [long-dashed green] and $0.60$ [medium-dashed blue], while results for the $1s$, $2s$ and $3s$ excitons are shown as horizontal gray lines.}
\label{fig:fig_pol_avgsep}
\end{center}
\end{figure*}

In Fig.~\ref{fig:fig_pol_avgsep}, we display the variation of the (root-mean-square) electron-hole separation as a function of detuning, for several different magnetic fields and light-matter coupling strengths.  We observe clear evidence of avoided crossings, which become more pronounced with increasing $\Omega/R$. The electron-hole separation within the polaritons generically increases monotonically with $\delta/R$. However, while the electron-hole separation in the LP is always smaller than that of the $1s$ exciton and mostly insensitive to the magnetic field, in the UP it is always larger, interpolating between the $1s$- and $2s$-exciton radii with increasing detuning. Most interestingly, we see both light- and magnetic-field-induced differences in the wave functions, thus highlighting the potential to engineer the matter excitations by using external fields.

\begin{figure}[h]
\begin{center}
\hspace{-3.2mm}
\includegraphics[trim = 23mm 2mm 29mm 21mm, clip, scale = 0.45]{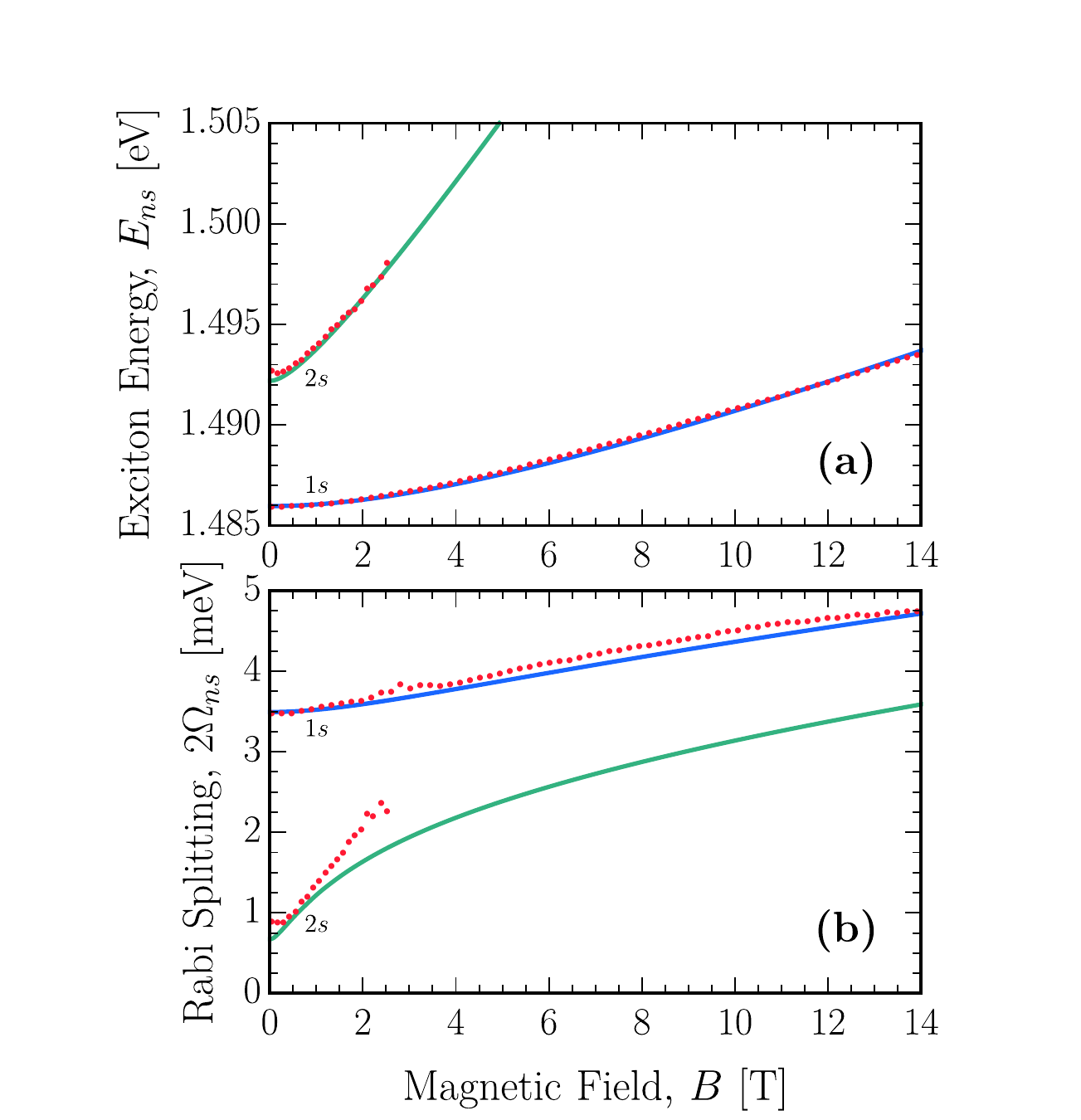}
\caption{Numerically exact exciton energies (a) and polariton Rabi splittings (b) for the $1s$ [blue solid lines] and $2s$ states [green solid lines], using the experimental parameters for a single GaAs quantum well embedded in a microcavity~\cite{PhysRevB.96.081402}.  The\linebreak red points correspond to the experimental data from Ref.~\cite{PhysRevB.96.081402}.}
\label{fig:fig_pol_Pietka}
\end{center}
\end{figure}

\begin{figure*}[t]
\begin{center}
\includegraphics[trim = 0mm 0mm 0mm 0mm, clip, scale = 1.675]{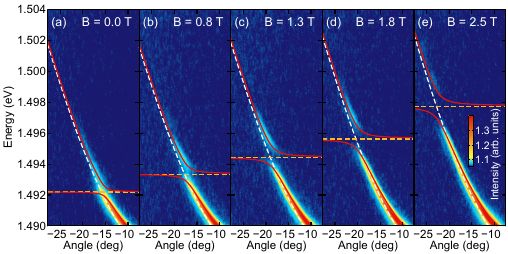}
\caption{Angle-resolved photoluminescence spectra of exciton-polaritons in a magnetic field taken from Ref.~\cite{PhysRevB.96.081402}, where our theory has been overlaid.  The white and orange dashed lines, respectively, correspond to the bare photon and $2s$-exciton dis-\linebreak persions, which anticross at high emission angles to form the two measured polariton branches.  The red solid lines are given by the eigenvalues of a $3$-level coupled oscillator model~\eqref{eq:COM} that is based on our numerically exact values for $E_{1s,\hspace{0.05em}2s}$ and $\Omega_{1s,\hspace{0.05em}2s}$ at the selected magnetic field strengths.  Full image credit:  Mateusz Kr\'{o}l.}
\label{fig:fig_Mateusz}
\end{center}
\end{figure*}

\section{Comparison to Experiment}
\label{sec:Theoretical_Comparison_to_Experiment}

We now demonstrate that the results of our microscopic theory compare well with two qualitatively different recent experiments \cite{PhysRevB.96.081402,PhysRevLett_119_027401} which explored the combination of strong light-matter coupling and strong magnetic fields.  In particular, Ref.~\cite{PhysRevB.96.081402} showed how the matter component of exciton-polaritons is modified by a very strong magnetic field. By contrast, in the experiment of Ref.~\cite{PhysRevLett_119_027401}, the use of a strong magnetic field enabled the observation of a non-perturbative modification of the exciton in the very strong light-matter coupling regime. Hence, both experiments illustrated the potential to directly engineer the fundamental matter excitations of a semiconductor microcavity.

\subsection{Polaritons in a strong magnetic field}
\label{sec:exp_strong}

In the experiment of Ref.~\cite{PhysRevB.96.081402}, the authors obtained the  photoluminescence spectrum of exciton-polaritons in a magnetic field using an optical microcavity where the embedded semiconducting active medium was a single 8-nm-thick $\mathrm{In}_{0.04}\mathrm{Ga}_{0.96}\mathrm{As}$ quantum well. The spectrum was fitted to a model of three coupled oscillators~\eqref{eq:COM}, where the $1s$- and $2s$-exciton and photon energies, and the corresponding Rabi splittings, were fitting parameters. The extracted energies of the bare matter excitations $E_{1s,\hspace{0.05em}2s}$ and the Rabi splittings $2\Omega_{1s,\hspace{0.05em}2s}$ could then be plotted as functions of the magnetic field. The latter were identified as the minimal  splittings between the polariton spectral lines around the $1s$ and $2s$ states as functions of an additional tuning parameter --- such as the detuning or photon momentum.

In Fig.~\ref{fig:fig_pol_Pietka}, we compare our results from Secs.~\ref{sec:Exciton_Problem} and~\ref{sec:Exciton-Polariton_Problem} to this experiment~\cite{PhysRevB.96.081402},  where as inputs to our theory, we employ the authors' reported $1s$-exciton binding energy $R=7~\mathrm{meV}$ and reduced mass $\mu=0.046~m_0$ (with $m_0$ the free electron mass)~\cite{PhysRevB.91.075309}. In particular, panel (a) shows exceptional agreement between the measured energies and the exciton energies $E_{ns}$ calculated from our Eq.~\eqref{eq:exciton_kappa} as a function of magnetic field $B$.  Note that we have applied a constant energy shift to the theoretical energies to match the quantum-well band gap.

In Fig.~\ref{fig:fig_pol_Pietka}(b), we overlay the measurements of the polariton Rabi splittings $2\Omega_{ns}$ onto our results, using the experimentally reported zero-field Rabi coupling strength, $\Omega=1.75~\mathrm{meV}$~\cite{PhysRevB.91.075309,PhysRevB.96.081402}.  Numerically, we can obtain the Rabi splittings in two different ways.  The first is by calculating the exciton wave functions at vanishing separation:  $\Omega_{ns}\hspace{-0.3mm}=\Omega\hspace{0.05em}\varphi_{ns}(0)/\varphi_{1s}^\mathrm{hyd}(0)$, as used in the COM~\eqref{eq:COM} and shown in Fig.~\ref{fig:fig_exciton_3}.  The second is by extracting $2\Omega_{ns}$ as the minimal polariton splittings when the energies of Eq.~\eqref{eq:phikappa_eqs_final} are plotted at fixed $B$ as functions of detuning.  In this experiment, the Rabi coupling is sufficiently weak ($\Omega/R\approx0.25$) that the COM with our numerically exact input parameters $\{E_{ns},\,\Omega_{ns}\}$ yields nearly exactly the same theoretical values as our full polariton calculation~\eqref{eq:phikappa_eqs_final}; by contrast, in Fig.~\ref{fig:fig_pol_COMs}, the failure of the COM is due to the calculation being in the strong light-matter coupling regime where $\Omega/R\gtrsim0.50$.  In panel (b) of Fig.~\ref{fig:fig_pol_Pietka}, we can see that the agreement between theory and experiment is very good for $\Omega_{1s}$.  However, the experimental value of $\Omega_{2s}$ deviates from our theory with increasing magnetic field, which is most likely due to the large uncertainty when fitting the polariton dispersion at low emission intensity~\footnote{B. Pietka, private communication.}.

To further validate our theory, in Fig.~\ref{fig:fig_Mateusz} we have overlaid its predictions directly onto the photoluminescence emission signals measured in this experiment~\cite{PhysRevB.96.081402}.  These experimental spectra clearly show the anticrossing of the bare cavity photon mode with the $2s$ exciton, as a function of the emission angle (which is proportional to the in-plane polariton momentum), at several magnetic field strengths.  The theoretical polariton dispersions have be-\linebreak en evaluated by diagonalizing a $3$-level coupled oscillator model~\eqref{eq:COM} based on the bare photon dispersion and our numerically exact $1s$ and $2s$ exciton energies and Rabi splittings (shown in Figs.~\ref{fig:fig_exciton_2} and~\ref{fig:fig_exciton_3}, respectively). 
Here, a change in the polariton's in-plane momentum simply corresponds to a change in the exciton-photon detuning, since the momentum probed is negligible compared with other relevant scales in the exciton.
We see that the calculated polariton modes fit the photoluminescence data perfectly at all fields considered, $B=0\hspace{0.15em}\text{--}\hspace{0.10em}2.5\mathrm{T}$.  Furthermore, this comparison makes it very clear that our theory consistently yields the correct minimal (Rabi) splittings between the polariton branches for the $2s$ state.

\subsection{Polaritons in the very strong light-matter coupling regime}

The regime of very strong light-matter coupling occurs when the Rabi frequency $\Omega$ becomes comparable to the $1s$-exciton binding energy $R$. In this limit, the light-matter coupling non-perturbatively modifies the electron-hole wave functions, effectively hybridizing the Rydberg series of excitons~\cite{KHURGIN2001307,Citrin2003} (also refer to the fully microscopic calculation of Ref.~\cite{PhysRevRes_1_033120}).

We now compare our theory to an experiment which demonstrated this effect by applying a magnetic field and determining the diamagnetic shifts of the upper and lower polaritons in a GaAs quantum-well microcavity --- see Ref.~\cite{PhysRevLett_119_027401}.  Measurements were performed on two different microcavity samples: one comprising a single quantum well (1~QW), and another comprising 28 quantum wells (28~QWs) placed in stacks of 4 quantum wells in the 7 central antinodes of the cavity light field.  All quantum wells were 7-nm-wide GaAs layers with 4-nm-wide AlAs barriers. Polariton energies and linewidths were obtained by fitting Lorentzian functions to the peaks observed in reflectance spectra. Note, the use of multiple quantum wells embedded into the microcavity allows one to increase the Rabi splitting roughly by a factor $\sim\!\sqrt{N}$, where $N$ is the quantum-well number.

\begin{figure*}[t]
\begin{center}
\includegraphics[trim = 7mm 25mm 9mm 26mm, clip, scale = 0.45]{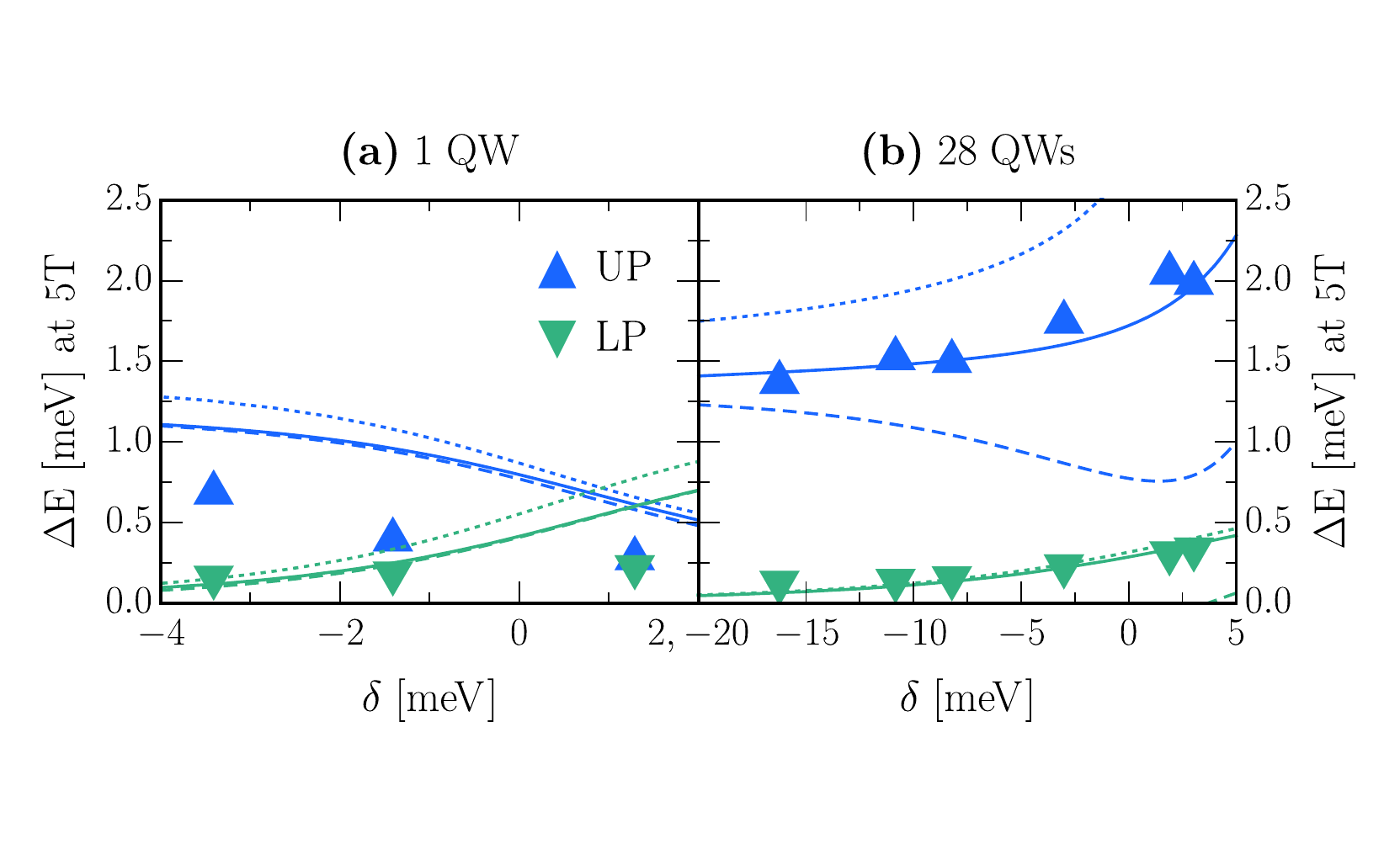}
\caption{Diamagnetic shifts of the upper and lower polaritons at $B=5\mathrm{T}$ as a function of zero-field detuning.  Results for the 1 quantum-well sample ($\Omega/R\approx0.14$) are shown in panel (a), while those for the 28 quantum-well sample ($\Omega/R\approx0.64$) are shown in panel (b).  The triangular markers correspond to the experimental data from Ref.~\cite{PhysRevLett_119_027401} and the solid lines correspond to our numerically exact theory.  The dashed lines are the results of a $2$-level COM~\eqref{eq:COM} in (a) and a $3$-level COM in (b), which treat the light-matter coupling as a perturbation.  On the other hand, the dotted lines in both panels are obtained by treating the magnetic field perturbatively.}
\label{fig:fig_pol_shifts}
\end{center}
\end{figure*}

Figure~\ref{fig:fig_pol_shifts} shows the diamagnetic shifts, $\Delta\mathrm{E}=E(B)-E(B=0)$, of the lower and upper polaritons at a magnetic field of $B=5\mathrm{T}$ ($\omega_c/R\approx1/2$) as a function of the zero-field detuning, $\delta$.  The triangular plot markers correspond to the experimental data taken from Ref.~\cite{PhysRevLett_119_027401}, while the solid lines correspond to our numerically exact theory~\eqref{eq:phikappa_eqs_final}.  The input parameters for the latter were the zero-field $1s$-exciton binding energy ($R=13.5~\mathrm{meV}$) and Rabi couplings ($\Omega=1.9~\mathrm{meV}$ for 1 QW and $\Omega=8.7~\mathrm{meV}$ for 28 QWs), as reported in Ref.~\cite{PhysRevLett_119_027401}.

For the 28-QW sample, the agreement between theory and experiment shown in panel (b) is quantitatively very close, which validates our microscopic calculation.  For 1 QW, the agreement shown in panel (a) is mainly qualitative; however the experimentalists state that there are considerable uncertainties in their measurements for this sample~\cite{PhysRevLett_119_027401}, possibly with a misestimate of the $1s$-exciton diamagnetic shift, $E_{1s}(B)-E_{1s}(B=0)$.  Moreover, the linewidth $\sim2\text{--}5~\mathrm{meV}$ is comparable to the Rabi splitting, and hence the 1-QW case is formally beyond the regime of validity of our strong-coupling theory.

To provide additional insight into the experimental results of Fig.~\ref{fig:fig_pol_shifts}, we compare them with two qualitatively distinct approximations: one which treats the coupling to light perturbatively, and one which instead assumes the magnetic field to be weak. In the former case, we calculate the numerically exact $1s$-exciton energy (as in Fig.~\ref{fig:fig_exciton_2}) and oscillator strength (as in Fig.~\ref{fig:fig_exciton_3}) using the methods of Sec.~\ref{sec:Exciton_Problem}, and then we obtain the polariton energies by using the simple coupled oscillator model in Eq.~\eqref{eq:COM}.  We see in panel (a) that the results obtained from a 2-level COM match our exact numerical results very well for 1~QW, as expected, since in this case $\Omega\ll R$. On the other hand, for 28~QWs we see in panel (b) that the perturbative results of a 3-level COM fail both quantitatively and qualitatively (in the considered detuning range) due to the very strong light-matter coupling, as discussed below.

Conversely, we can first obtain the LP and UP energies as the two lowest energy eigenstates of Eq.~\eqref{eq:phikappa_eqs_final} with $B=0$, and then add the magnetic field as a perturbation, giving
\begin{align}\label{eq:diamag}
    \Delta \mathrm{E}_{{\tiny  \begin{matrix} \text{LP}\\\text{UP}\end{matrix}}}\simeq \frac12 \mu\omega_c^2 \left<r^2\right>_{{\tiny  \begin{matrix} \text{LP}\\\text{UP}\end{matrix}}} =\frac12 \mu\omega_c^2(1-|\gamma|^2) \left<r^2\right>_{{\hspace{-0.50mm}\varphi,\hspace{+0.25mm}\tiny  \begin{matrix} \text{LP}\\\text{UP}\end{matrix}}} ,
\end{align}
where $\left<r^2\right>_{{\hspace{-0.50mm}\varphi}}$ is calculated from the matter-only part of the polariton state as in Eq.~\eqref{eq:r_squared}.  Qualitatively, this approach captures the behavior of both LP and UP diamagnetic shifts with detuning in both samples. In particular, for 28~QWs it works very well for the lower polariton, due to this being dominated by the coupling to the $1s$ exciton, which is only minimally affected by this magnitude of magnetic field.  On the other hand, the very strong light-matter coupling means that the upper polariton has contributions from multiple Rydberg states as well as the continuum~\cite{KHURGIN2001307,Citrin2003,PhysRevRes_1_033120}, which are strongly affected by the magnetic field, and hence this approach overestimates the diamagnetic shift.  Together, the results from these two perturbative calculations clearly demonstrate that the measurements for the 28-QW sample can only be accurately described by a theory that incorporates both very strong light-matter coupling and strong magnetic fields.

\begin{figure*}[t]
\begin{center}
\includegraphics[trim = 7mm 25mm 9mm 26mm, clip, scale = 0.45]{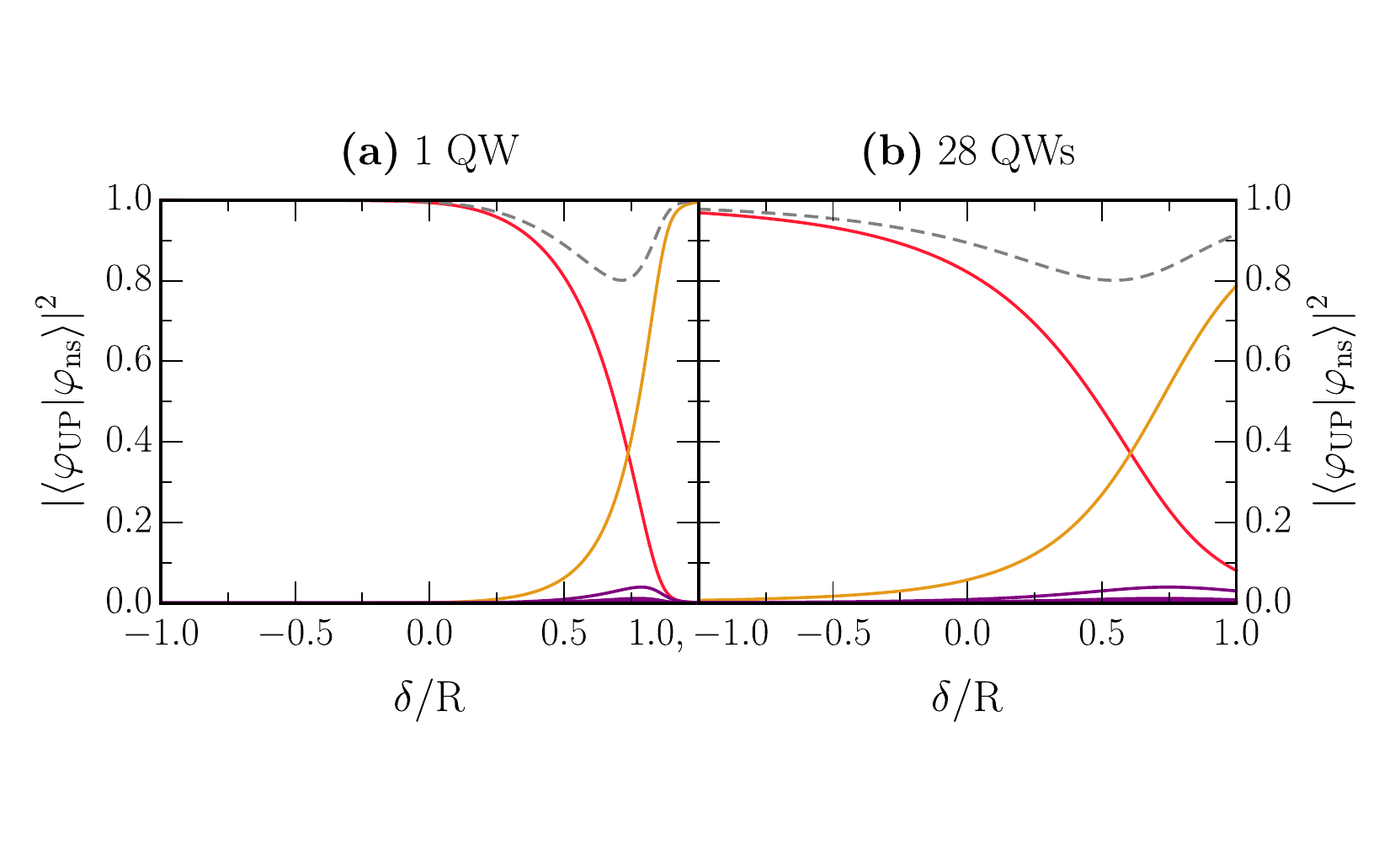}
\caption{Exciton fractions of the upper polariton at $B=0$ as a function of detuning, and for the same light-matter coupling parameters as in Fig.~\ref{fig:fig_pol_shifts}.  In each panel, the solid red, orange and purple lines respectively correspond to the $1s$, $2s$ and $3s\hspace{-0.33mm}-\hspace{-0.68mm}10s$ excitonic states, while the dashed gray line represents the total fraction of all the bound states.}
\label{fig:fig_pol_overlaps}
\end{center}
\end{figure*}

As evidenced by Eq.~\eqref{eq:diamag}, the qualitative behavior of our numerically exact results arises from the interplay between the exciton fraction $(1-|\gamma|^2)$ and the electron-hole (matter-only) separation [see Fig.~\ref{fig:fig_pol_avgsep}(a)--(b)].  In particular, since both of these quantities for the lower polariton increase with increasing detuning, the LP diamagnetic shift always monotonically increases with $\delta/R$ for any Rabi coupling.  However, this shift becomes suppressed with increasing $\Omega/R$ due to the light-induced shrinkage of the exciton.  We remark that an $(n+1)$-level COM cannot capture this effect and instead produces an unphysical negative LP diamagnetic shift --- see, e.g., Fig.~\ref{fig:fig_pol_shifts}(b) where $\Delta\mathrm{E}_\text{LP}$ as predicted by the COM is negative in almost the entire detuning range.  In other words, it is thus essential to include the back-action of light on matter in order to correctly capture the polariton diamagnetic shift.

By contrast, the $1s$-exciton fraction in the upper polariton decreases with increasing detuning, and thus the diamagnetic shift in Eq.~\eqref{eq:diamag} depends on the competition between this effect and the expanding size of the exciton.  For the sufficiently small Rabi coupling in Fig.~\ref{fig:fig_pol_shifts}(a), we see that the overall product of these two factors decreases, which means that the effect of the changing exciton fraction is dominant.  The result is that the LP and UP shifts behave symmetrically as a function of detuning in this limit, coinciding at $[E_{1s}(B)-E_{1s}(B=0)]/2$ where they both possess the same photon fraction.  As the detuning increases beyond the range shown in panel (a), the UP becomes essentially indistinguishable from the $2s$ exciton, and therefore the $2s$-exciton fraction approaches unity, while $\Delta\mathrm{E}_{\text{UP}}$ turns upward and approa-\linebreak ches $E_{2s}(B)-E_{2s}(B=0)$.  As $\Omega/R$ increases between the values in panels (a) and (b), this upturn occurs at increasingly smaller values of $\delta/R$.  At the strong Rabi coupling considered in Fig.~\ref{fig:fig_pol_shifts}(b), a 3-level COM which includes this phenomenology completely fails to describe the experimental and exact theoretical results, and including more levels in the COM does not lead to an improved comparison~\footnote{Note that at finite $B$ fields, the resonant cavity photon energy within the $(n+1)$-level COM formally diverges in the limit $n\to\infty$, analogous to the divergence in Eq.~\eqref{eq:detuning}.}.  Instead the perturbative COM becomes an even worse predictor because it cannot account for the light-induced changes to the matter part of the polariton in this regime.

The importance of the admixture of exciton Rydberg states in the UP is illustrated in Fig.~\ref{fig:fig_pol_overlaps}. Here, we present the exciton fractions of the upper polariton for the same Rabi coupling strengths as in panels (a) and (b) of Fig.~\ref{fig:fig_pol_shifts} and for a fixed detuning range --- in the limit of zero magnetic field.  This was done by calculating the overlap of the exciton wave function with the electron-hole part of the polariton wave function (both normalized to unity).  The colored lines correspond to the first ten exciton states, and the dashed gray line represents the sum total of these contributions for all the bound states. After the $1s$ exciton (in red), the $2s$ exciton (in orange) has the next most dominant effect on the behavior of the UP.  However, at large Rabi coupling and large detuning, there is an additional effect from higher energy excitonic states ($3s$ up to $10s$ are shown in purple) and from the continuum (since the gray line can be less than one).  Notice that the larger the value of $\Omega$, the smaller the value of $\delta$ at which these states become important.  This effectively means that we cannot explain the very large shift of the UP with a coupled oscillator model that includes the photon and the $1s$, $2s$, or even higher $ns$ excitons.  Instead, we need to microscopically consider all the bound and unbound electron-hole levels in order to quantitatively capture the very strong light-matter coupling regime.

\section{Conclusions and Outlook}
\label{sec:Conclusions_and_Outlook}

To conclude, we have developed a microscopic theory of Rydberg exciton-polaritons in the presence of a transverse magnetic field. Our theory is underpinned by two key innovations: a mapping between the 2D hydrogen problem and the 2D harmonic oscillator that allows a highly efficient numerical solution of the entire $s$-exciton Rydberg series in a magnetic field; and a recently developed renormalization scheme that enables the numerically exact description of exciton-polaritons~\cite{PhysRevRes_1_033120}. As the first such theory, it can be utilized to determine the effect of tuning both the light-matter coupling and the magnetic field to arbitrary values, and in particular to investigate the interplay between very strong light-matter coupling and strong magnetic fields.

We emphasize that the excellent agreement between our theory and the two experiments in Refs.~\cite{PhysRevB.96.081402,PhysRevLett_119_027401} was achieved without the need to introduce any fitting parameters. Rather, all parameters used in our numerically exact modeling were taken from independent experimental measurements. Furthermore, we have modeled the experiments as purely two-dimensional, and our agreement with the experimental results therefore suggests that the transverse motion in the quantum wells plays only a minor role.  In the future, it would be interesting to apply the theory to other semiconductors such as ${\mathrm{Cu}}_{2}\mathrm{O}$~\cite{Ziemkiewicz_PRB2021}.

Our results have a wide regime of applicability: to any electron-hole mass ratio, to any polariton momentum below the inflection point, and to different semiconductor structures or materials.  In particular, our approach is immediately extendable to study Rydberg exciton-po-\linebreak laritons in TMD monolayers, since the excited excitons in these systems are already well captured by the Coulomb potential considered in this work (see Appendix~\ref{sec:Appendix_B}).  This direction is especially promising now that experiments have demonstrated how large magnetic fields in TMDs can be induced by strain~\cite{C2CP42181J,strain_TMDs}.

One particularly exciting aspect of Rydberg exciton-polaritons is their potential as strongly interacting part light, part matter quasiparticles, with applications to quantum information processing~\cite{Walther2018}. Indeed, enhanced interactions of the $2s$ Rydberg exciton-polaritons have very recently been reported in an atomically thin semiconductor~\cite{Gu2021}. Furthermore, polariton scattering calculations that explicitly included the electron, hole, and photon degrees of freedom have recently demonstrated that the interactions in light-matter coupled systems can be enhanced even above that of the matter-only excitations~\cite{Bleu2020,li2021microscopic}, and it would be interesting to extend such microscopic approaches to the scattering of Rydberg exciton-polaritons in magnetic fields.  Our work elucidates how Rydberg excitons can be strongly modified both by light-matter coupling and via a magnetic field, thus highlighting the prospect of precisely engineering Rydberg states with enhanced light-matter coupling and strong non-linearities.

\acknowledgements

We thank Christian Schneider and Barbara Pietka for useful discussions, Sebastian Brodbeck and Christian Schneider for sharing the data from Ref.~\cite{PhysRevLett_119_027401}, and Mateusz Kr\'{o}l and Barbara Pietka for sharing the data from Ref.~\cite{PhysRevB.96.081402}.  In addition, we are grateful to Mateusz Kr\'{o}l for generating Fig.~\ref{fig:fig_Mateusz} and performing the associated coupled oscillator model calculation.  This work was supported by the Australian Research Council Center of Excellence in Future Low-Energy Electronics and Technologies (CE170100039).  E.L. was supported by a Graduate Research Completion Award and by a Postgraduate Publications Award at Monash University.  M.M.P. and J.L. were supported by the Australian Research Council Future Fellowships, FT200100619 and FT160100244, respectively.  F.M.M. acknowledges financial support from the Ministerio de Ciencia e Innovaci\'on (MICINN) project No.~AEI/10.13039/501100011033 (2DEnLight), as well as from Proyecto Sinérgico CAM 2020 Y2020/TCS-6545 (NanoQuCo-CM).

\appendix

\section{Subtraction Scheme}
\label{sec:Appendix_A}

Here, we briefly illustrate the idea behind the subtraction scheme employed to deal with the pole at $\bm\kappa=\bm\kappa'$ in the two off-diagonal terms of Eqs.~\eqref{eq:exciton_kappa} and~\eqref{eq:phikappa_eqs_final}, and to speed up the numerical convergence.  Let us consider the simple Coulomb potential as an example;  the same procedure can be readily generalized to any potential  of the form $\beta|\bm\kappa-\bm\kappa'|^{\alpha}$ with a pole at $\bm\kappa=\bm\kappa'$ for $-1\le\alpha<0$. In dimensionless units, the Coulomb eigenvalue problem is
\begin{align}
\label{eq:eigen_simple}
\kappa^2-\int{d}\kappa'\kappa'\hspace{0.3mm}W_{\kappa,\hspace{0.1mm}\kappa'}\varphi_{\kappa'}=\bar{E}\varphi_{\kappa}\,,
\end{align}
where we have integrated over the angle,
\begin{align}
W_{\kappa,\hspace{0.1mm}\kappa'}=\int_0^{2\pi}\frac{d\theta}{2\pi}\frac{1}{|\bm\kappa-\bm\kappa'|}=\frac{2}{\pi}\frac{1}{\kappa+\kappa'}\mathrm{K}\!\left[\frac{4\kappa\kappa'}{(\kappa+\kappa')^2}\right]\!,
\end{align}
and where $\mathrm{K}(x)$ denotes the complete elliptic integral of the first kind.  Problematically, the potential $W_{\kappa,\hspace{0.1mm}\kappa'}$ diver-\linebreak ges at $\kappa=\kappa'$.  We therefore consider the following equivalent eigenvalue problem:
\begin{align}
\label{eq:subtraction}
&\kappa^2-\int{d}\kappa'\kappa'\big(W_{\kappa,\hspace{0.1mm}\kappa'}\varphi_{\kappa'}-g_{\kappa,\hspace{0.1mm}\kappa'}\varphi_{\kappa}\big)-\int{d}\kappa'\kappa'g_{\kappa,\hspace{0.1mm}\kappa'}\varphi_{\kappa}\nn\\&\hspace{-1.0mm}=\bar{E}\varphi_{\kappa}\,,\quad\text{with}\hspace{1.3mm}{g}_{\kappa,\hspace{0.1mm}\kappa'}=\frac{2\kappa^2}{\kappa^2+(\kappa')^2}W_{\kappa,\hspace{0.1mm}\kappa'}\,.
\end{align}
Above, we have added and subtracted the potential $g_{\kappa,\hspace{0.1mm}\kappa'}$ in the diagonal term of the original eigenvalue problem \eqref{eq:eigen_simple}.  The prefactor $2\kappa^2/\big[\kappa^2+(\kappa')^2\big]$ has been chosen so that it equals $1$ when $\kappa\to\kappa'$, and it increases the convergence rate when $\kappa'\to\infty$~\cite{PhysRevB_43_1500}.  The singularity of the original potential is thus removed in the new eigenvalue problem~\eqref{eq:subtraction}, while the last term on the left-hand side is convergent.  For the bare Coulomb potential, this term can be evaluated analytically, giving
\begin{equation}
\int{d}\kappa'\kappa'g_{\kappa,\hspace{0.1mm}\kappa'}=\frac{\kappa}{2\sqrt{2}}\frac{\Gamma^2(1/4)}{\Gamma(1/2)}\,.
\end{equation}

\section{Transformations for a Generic Electron-Hole Potential}
\label{sec:Appendix_B}

Our approach can be generalized to different electron-hole interaction potentials, and we illustrate this below on the solution for the exciton, which is contained within the solution for the polariton.  In particular, we confirm that this generalization works for the Rytova--Keldysh potential~\cite{Rytova,Keldysh,PhysRevB_84_085406}, which properly captures the dielectric screening in monolayer TMDs~\cite{PhysRevB_89_205436,PhysRevLett_113_076802}.

In momentum space, this potential has the following form:
\begin{align}
\label{eq:k-space_Keldysh}
V^\mathrm{RK}_k=V^\mathrm{C}_k\!\left(\frac1{1+r_0k}\right),\quad\mathrm{where}\quad{V}^\mathrm{C}_k=-\frac{2\pi{e}^2}{\varepsilon{k}}
\end{align}
is the 2D Coulomb potential [i.e., the Fourier transform of Eq.~\eqref{eq:Coulomb_potential}], and $k$ is the magnitude of the relative momentum of the electron and hole.  The ``screening length'' $r_0$ provides a correction to the Coulomb potential which is clearly only important at large momenta, or short distances.  As a consequence, the Rytova--Keldysh potential only strongly affects the tightest bound excitons in TMD monolayers, while the excited states in the Rydberg series remain well described by $V^\mathrm{C}_k$.  In position space, the Fourier transform of $V^\mathrm{RK}_k$ is given by
\begin{align}
\label{eq:r-space_Keldysh}
V^\mathrm{RK}(r)=-\frac{\pi{e}^2}{2\varepsilon{r_0}}\left[\mathrm{H}_0\!\left(\frac{r}{r_0}\right)-\mathrm{Y}_0\!\left(\frac{r}{r_0}\right)\right],
\end{align}
where the functions $\mathrm{H}_0$ and $\mathrm{Y}_0$ have been defined below Eq.~\eqref{eq:phikappa_eqs_final}.

We write the real-space Schr\"{o}dinger equation for the exciton~\eqref{eq:exc_s-wave_with-B_Ch7}, from the main text, in terms of a generic rotationally symmetric potential, $V(\r)\equiv{V}(r)$:
\begin{align}
\label{eq:SE_with_generic_potential}
E\varphi(r)=\left[-\frac1{2\mu}\!\left(\frac{d^2}{dr^2}+\frac1r\frac{d}{dr}\right)\!+\frac{\mu\omega_c^2}2r^2+V(r)\right]\!\varphi(r)\,.
\end{align}
Continuing, we rescale the real-space variable as $r\to\rho=\linebreak{r}^2/(8a_0^2)$, and we also multiply the entire expression by $2a_0/(R\rho)$ to yield
\vspace{-0.7mm}
\begin{align}
\label{eq:rho-eq_with_gen-pot}
\hspace{-3.9mm}\frac{2\hspace{0.01em}\bar{E}}\rho\hspace{0.1em}\widetilde\varphi(\rho)=\left[-\!\left(\frac{d^2}{d\rho^2}+\frac1\rho\frac{d}{d\rho}\right)\!+4\hspace{0.1em}\bar\omega_c^2+\widetilde{V}(\rho)\right]\!\widetilde\varphi(\rho)\,,
\end{align}
where $\widetilde\varphi(\rho)=a_0\varphi(r)$.  Notice that we have absorbed the multiplicative factor $2/(R\rho)$ into the generic potential,
\begin{align}
\label{eq:gen_pot}
\widetilde{V}(\rho)\equiv\frac{2V\big(\sqrt{8a_0^2\rho}\,\big)}{R\rho}\,.
\end{align}
Next, we define a rescaled conjugate variable $\kappa$, and we Fourier transform the equation into momentum space by acting on both sides with the operator $\int{d}^{\hspace{0.05em}2}\!\hspace{0.10em}\bm\rho\,{e}^{-i\bm\kappa\cdot\bm\rho}\{\,\cdot\,\}$.  In particular, the potential term transforms as follows:
\begin{align}
\label{eq:FT_gen-pot}
&\int{d}^2\!\bm\rho\,e^{-i\bm\kappa\cdot\bm\rho}\widetilde{V}(\rho)\widetilde{\varphi}(\rho)\nn\\&\hspace{-0.4mm}=\int{d}^2\!\bm\rho\,e^{-i\bm\kappa\cdot\bm\rho}\sum_{\bm\kappa''}e^{i\bm\kappa''\cdot\bm\rho}\widetilde{V}_{\kappa''}\sum_{\bm\kappa'}e^{i\bm\kappa'\cdot\bm\rho}\widetilde{\varphi}_{\kappa'}\nn\\&\hspace{-0.4mm}=\sum_{\bm\kappa'}\widetilde{V}_{|\bm\kappa\hspace{0.1mm}-\hspace{0.1mm}\bm\kappa'|}\widetilde{\varphi}_{\kappa'}\,,
\end{align}
where we have used the Dirac delta function $\delta^2(\bm\kappa'+\bm\kappa''\hspace{-0.6mm}-\bm\kappa)=\int{d}^{\hspace{0.05em}2}\!\hspace{0.10em}\bm\rho\,e^{i(\bm\kappa'\hspace{-0.1mm}+\,\bm\kappa''\hspace{-0.2mm}-\,\bm\kappa)\cdot\bm\rho}$, and where $\sum_{\bm\kappa}\equiv\int{d}^{\hspace{0.05em}2}\!\hspace{0.10em}\bm\kappa/(2\pi)^2$.  The final result in the rescaled momentum space is thus
\begin{align}
\label{eq:kappa-eq_with_gen-pot}
\hspace{-2.2mm}\bar{E}\sum_{\bm\kappa'}\frac{4\pi\hspace{0.1em}\widetilde{\varphi}_{\kappa'}}{|\bm\kappa-\bm\kappa'|}=\kappa^2\hspace{0.1em}\widetilde{\varphi}_{\kappa}+4\hspace{0.1em}\bar{\omega}_c^2\hspace{0.1em}\widetilde{\varphi}_{\kappa}+\sum_{\bm\kappa'}\widetilde{V}_{|\bm\kappa\hspace{0.1mm}-\hspace{0.1mm}\bm\kappa'|}\hspace{0.1em}\widetilde{\varphi}_{\kappa'}\,.
\end{align}

Importantly, we can carry out the Fourier transform from $\bm\rho$ to $\bm\kappa$ space as long as the original real-space potential does not diverge too strongly at short range and decays at long range.  To be precise, at small $\rho$ we require that $\widetilde{V}(\rho)\sim\rho^\alpha$ where $\alpha>-2$, which means in turn that we need to have $V(r)\sim{r}^\beta$ at small $r$, with $\beta=2\alpha+2\linebreak>-2$.  This condition is satisfied by the Rytova--Keldysh potential $V^\mathrm{RK}(r)$, since that potential goes as $\mathrm{ln}(r)$ for $r\to0$, which is a weaker divergence than $r^{-2}$.  In the opposite limit we simply require that $V(r)\to0$ for $r\to\infty$, and this is also satisfied by $V^\mathrm{RK}(r)$, which reduces to the Coulomb potential $V^\mathrm{C}(r)\propto{r}^{-1}$ at large $r$.

\bibliography{mag_refs.bib}

\end{document}